\documentclass{article}
\usepackage{amsmath}
\usepackage{amsfonts}
\usepackage{amssymb}
\usepackage{latexsym,epic,eepic}
\begin{document}
\begin{flushright}
HUPD-9916\\
September 1999
\end{flushright}
\vspace{5mm}
\begin{center}
{\Large \bf
Structure of Chiral Phase Transitions at Finite Temperature\\
in Abelian Gauge Theories}
\\[8mm]
{\large Kenji Fukazawa}\\
Kure National College of Technology, Kure, Hiroshima 737-
8506, Japan\\[2mm]
{\large Tomohiro Inagaki}\\
Information Processing Center, Hiroshima University, 
Higashi-Hiroshima,\\
Hiroshima 739-8526, Japan\\[2mm]
{\large Seiji Mukaigawa}\\
Research Center for Nanodevices and Systems, Hiroshima 
University,\\
Higashi-Hiroshima, Hiroshima 739-8527, Japan\\[2mm]
{\large Taizo Muta}\\
Department of Physics, Hiroshima University,
Higashi-Hiroshima,\\
Hiroshima 739-8526, Japan\\[16mm]
\end{center}

\begin{abstract}

     The mechanism of the chiral symmetry breaking is investigated
in the strong-coupling Abelian gauge theories at finite temperature.
The Schwinger-Dyson equation in Landau gauge is employed
in the real time formalism and is solved numerically within the
framework of the instantaneous exchange approximation including the
effect of the hard thermal loop for the photon propagator.
It is found that the chiral symmetry is broken below the critical
temperature $T$ for sufficiently large coupling $\alpha$.
The chiral phase transition is found to be of the 2nd order
and the phase diagram on the $T-\alpha$ plane is obtained.
It is investigated how the structure of the chiral phase transition
is affected by the hard thermal loops in the photon propagator.

\end{abstract}




\newpage

\section{Introduction}
     The chiral phase transition in gauge field theories 
at finite temperature is an interesting phenomenon 
in many respects. 
In particular it plays an important role when we deal 
with the early stage of the universe.
     It has been known for a long time that Abelian 
gauge theories at vanishing temperature subject to the 
chiral symmetry breaking in the strong coupling region 
and hence the massless fermions acquire a mass in a 
dynamical way \cite{BJW,MN,FK,Mir}.
This phenomenon has been clarified mostly by using 
the Schwinger-Dyson equation \cite{SD}.

It is quite natural from the point of view of the early 
universe to introduce the temperature effect in the 
analysis of the above phenomenon and see whether the 
broken chiral symmetry at vanishing temperature is
restored at sufficiently high temperature and examine 
whether the phase transition is of the first order or 
second order. There have been several works dealing with 
this problem in quantum chromodynamics. There have, 
however, been not many works \cite{Aki,KY,Tri} which 
have studied this question by taking into account of 
the photon mass coming from the hard thermal loop. Thus 
we try to present the thorough analysis of the chiral 
phase transition in Abelian gauge theories at finite
temperature with due consideration on the hard thermal 
loop by using the Schwinger-Dyson equation.

In our analysis we deal with the Schwinger-Dyson equation 
in the real time formalism to introduce the 
temperature \cite{UMT,Kap,LeB,Das}.
To solve the Schwinger-Dyson equation we confine 
ourselves to the instantaneous exchange approximation. 
The approximation is found to be valid in the high 
temperature region. In the case of the vanishing temperature 
the vacuum polarization effect is negligible. For high 
temperature the vacuum polarization becomes
enhanced and so within our approximation we are forced 
to take into account of the vacuum polarization function 
in the photon propagator while dealing with the 
Schwinger-Dyson equation.

We make mostly numerical calculations in our analysis 
of the finite-temperature Schwinger-Dyson equation. 
We find that our numerical solutions are quite stable and 
the results are proven to be reliable.
We find the clear signal of the second order chiral 
phase transition as temperature varies. 
We investigate the thermal photon mass effect in the 
chiral phase transition. We obtain critical curves on 
the $T-\alpha$ plane with $T$ the temperature 
and $\alpha$ the fine structure constant.

\section{Schwinger-Dyson equation at finite Temperature}

     Throughout the paper we work in Abelian gauge theories, in 
particular quantum electrodynamics, with massless fermions.
The main purpose of our work is to see whether the broken chiral
symmetry for large $\alpha$ at vanishing temperature is restored
at high temperature and, if so, what is the nature of the chiral
phase transition, i. e. whether the transition is of the 1st order
or the 2nd order. Here $\alpha$ is the fine structure constant
$\alpha=e^2/4\pi$ with $e$ the electric charge of fermions.
We rely on the Schwinger-Dyson equation in the real time formalism
of the finite-temperature field theory.
In the imaginary time approach the integral equation is cast into 
the infinite-component simultaneous equation and we do not prefer 
to cut off the summation in our calculation.

     We start with the brief summary on the Schwinger-Dyson equation
for vanishing temperature $T=0$. The Schwinger-Dyson equation for
the fermion self-energy part $\Sigma(p)$ reads within the ladder 
approximation in Landau gauge
\begin{equation}
   \Sigma(p) = -ie^2\int\frac{d^4q}{(2\pi)^4}
               \gamma^{\mu}iS(q)\gamma^{\nu}iD^{tree}_{\mu\nu}(p-q),
\label{T0SD}
\end{equation}
where $D^{tree}_{\mu\nu}(p-q)$ is the photon propagator at tree level
and the self-energy part $\Sigma(p)$ is defined through 
\begin{equation}
   iS(p)=\frac{i}{p\negthickspace /-\Sigma(p)+i\varepsilon}
        =\frac{i}{A(p^2)p\negthickspace /-B(p^2)+i\varepsilon},
\end{equation}
with $S(p)$ the full propagator for massless fermions, and
$A(p^2)$ and $B(p^2)$ the invariant functions of $p^2$ respectively.
By the use of the invariant functions the self-energy part is represented
such that $\Sigma(p)=(1-A(p^2))p\negthickspace /+B(p^2)$.
It has been known for a long time \cite{BJW,MN,FK} that Eq. (\ref{T0SD})
allows a non-vanishing solution for $B(p^2)$ with $A(p^2)=1$
if $\alpha > \alpha_{c}=\pi/3$.
Thus the chiral symmetry is broken for the unusually large 
electromagnetic coupling and the massless fermion acquires the
dynamical mass for $\alpha > \pi/3$.

     At finite temperature in the real time formalism 
\cite{UMT,Kap,LeB,Das} the full propagator for fermions 
may be written in the following form:
\begin{equation}
   iS(p)=\frac{i}{p\negthickspace /-\Sigma(p)+i\varepsilon}
   =\frac{i}{A_0(p_0,|\vec{p}|)p_0\gamma^0
    +A(p_0,|\vec{p}|)p_i\gamma^i-B(p_0,|\vec{p}|)+i\varepsilon}.
\end{equation}
For simplicity we assume that 
\begin{equation}
   A_0(p_0,|\vec{p}|)=A(p_0,|\vec{p}|)=1 , \ \ \
   \mbox{Im} B(p_0,|\vec{p}|)=0 .
\end{equation}
It will be seen that the above simplifying assumption does not lead
to any inconsistencies. We then have
\begin{equation}
   iS(p)=\frac{i}{p\negthickspace /-M(p_0,|\vec{p}|)+i\varepsilon} ,
\end{equation}
where the mass function $M(p_0,|\vec{p}|)$ is defined by
\begin{equation}
   M(p_0,|\vec{p}|)=\frac{\mbox{tr}{\mbox{Re} \Sigma(p)}}{\mbox{tr}1} .
\end{equation}
In the closed time path method \cite{Das} the spinor self-energy part 
is given by the following matrix form:
\begin{equation}
   i\Sigma^{ab}(p)=V^{-1}(\beta,p)
   \left(
   \begin{array}{cc}
   i\Sigma(p) & 0 \\
   0 & -i\Sigma^{*}(p)
   \end{array}
   \right)
   V^{-1}(\beta,p) ,
\end{equation}
where $V(\beta,p)$ is the unitary matrix which connects the thermal 
vacuum to the zero-temperature vacuum and is given by
\begin{equation}
   V(\beta,p)=
   \left(
   \begin{array}{cc}
   \cos\varphi & -\epsilon(p_0)\sin\varphi \\
   \epsilon(p_0)\sin\varphi & \cos\varphi
   \end{array}
   \right) ,
\end{equation}
with
\begin{equation}
   \cos\varphi=\frac{1}{\sqrt{1+\exp(-\beta|p_0|)}}; \ \ \
   \sin\varphi=\frac{\exp(-\beta|p_0|/2)}{\sqrt{1+\exp(-\beta|p_0|)}}.
\end{equation}
Accordingly we find
\begin{equation}
   \mbox{Re}\Sigma^{11}(p)=\mbox{Re}\Sigma(p),
\end{equation}
and hence
\begin{equation}
   M(p_0,|\vec{p}|)=\frac{\mbox{tr}{\mbox{Re} \Sigma^{11}(p)}}{\mbox{tr}1} .
\end{equation}

     The Schwinger-Dyson equation in the real time formalism is written 
down in the matrix form within the framework of the closed time path
method. If the vertex part is approximated by the tree form,
the 1-1 matrix element of the equation is represented only by the 1-1
component of the fermion and photon propagator respectively
and takes the following form,
\begin{equation}
   \Sigma^{11}(p)=-ie^2\int\frac{d^4q}{(2\pi)^4}
   \gamma^{\mu}iS^{11}(q)\gamma^{\nu}iD^{11}_{\mu\nu}(p-q).
\end{equation}
Here the spinor two-point function $S^{11}(p)$ is given by evaluating 
the 1-1 matrix element of the expression \cite{Aki,KY}
\begin{equation}
   iS^{ab}(p)=V(\beta,p)
   \left(
   \begin{array}{cc}
   S(p) & 0 \\
   0 & S^{*}(p)
   \end{array}
   \right)
   V(\beta,p) ,
\end{equation}
and reads
\begin{equation}
   iS^{11}(p)=(p\negthickspace /+M(p_0,|\vec{p}|))
   \left[\mathcal{P} \frac{i}{p^2-M^2(p_0,|\vec{p}|)}
   +\pi\delta(p^2-M^2(p_0,|\vec{p}|))\tanh\frac{\beta |p_0|}{2}\right].
\end{equation}
The photon two-point function $D^{11}_{\mu\nu}(p)$ is also given by
evaluating the 1-1 matrix element of the matrix
\begin{equation}
   iD_{\beta}^{ab}(q)=U(\beta,q)
   \left(
   \begin{array}{cc}
   D_{\mu\nu}(q) & 0 \\
   0 & D_{\mu\nu}^{*}(q)
   \end{array}
   \right)
   U(\beta,q) ,
\end{equation}
where
\begin{equation}
   U(\beta,q)=
   \left(
   \begin{array}{cc}
   \cosh\theta & \sinh\theta \\
   \sinh\theta & \cosh\theta
   \end{array}
   \right) ,
\end{equation}
\begin{equation}
   \cosh\theta=\frac{1}{\sqrt{1-\exp(-\beta|q_0|)}}; \ \ \
   \sinh\theta=\frac{\exp(-\beta|q_0|/2)}{\sqrt{1-\exp(-\beta|q_0|)}} .
\end{equation}
The general form of the photon propagator at finite temperature is
well-known and is given by \cite{LeB}
\begin{equation}
   iD_{\mu\nu}(q) = \frac{i}{q^2-\Pi^T(q)+i\varepsilon}P_{\mu\nu}^T
                 +\frac{i}{q^2-\Pi^L(q)+i\varepsilon}P_{\mu\nu}^L
                 -i\frac{\alpha_{GF}}{q^2+i\varepsilon}\frac{q_{\mu}q_{\nu}}
                 {q^2} ,
\end{equation}
where $P_{\mu\nu}^T$ and $P_{\mu\nu}^L$ are the transverse and longitudinal 
projection operators respectively and $\alpha_{GF}$ is the gauge fixing 
parameter. Below we take the Landau gauge $\alpha_{GF}=0$ which is 
consistent with the Ward-Takahashi identity within the ladder approximation 
at $T=0$.
We then find that
\begin{equation}
   iD^{11}_{\mu\nu}(p)=\mbox{Re} D_{\mu\nu}(q)\coth\frac{\beta|q_0|}{2}
                       +i\mbox{Im} D_{\mu\nu}(q),
\end{equation}
where
\begin{eqnarray}
   \mbox{Re} D_{\mu\nu}(q)&=&\mbox{Re} D^T(q)P_{\mu\nu}^T
             +\mbox{Re} D^L(q)P_{\mu\nu}^L , \\
   \mbox{Im} D_{\mu\nu}(q)&=& \mbox{Im} D^T(q)P_{\mu\nu}^T
             +\mbox{Im} D^L(q)P_{\mu\nu}^L .
\end{eqnarray}
with
\begin{eqnarray}
   \mbox{Re} D^T(q)&=& \mathcal{P}
   \frac{-\mbox{Im}\Pi^T(q)}{(q^2-\Pi^T(q))(q^2-{\Pi^T}^{*}(q))}
   \nonumber \\
   &&+\pi\varepsilon(q^2-\mbox{Re}\Pi^T(q))(q^2-\mbox{Re}\Pi^T(q))
   \delta((q^2-\Pi^T(q))(q^2-{\Pi^T}^{*}(q))),\\
   \mbox{Re} D^L(q)&=& \mathcal{P}
   \frac{-\mbox{Im}\Pi^L(q)}{(q^2-\Pi^L(q))(q^2-{\Pi^L}^{*}(q))}
   \nonumber \\
   &&+\pi\varepsilon(q^2-\mbox{Re}\Pi^L(q))(q^2-\mbox{Re}\Pi^L(q))
   \delta((q^2-\Pi^L(q))(q^2-{\Pi^L}^{*}(q))),\\
   \mbox{Im} D^T(q)&=& \mathcal{P}
   \frac{q^2-\mbox{Re}\Pi^T(q)}{(q^2-\Pi^T(q))(q^2-{\Pi^T}^{*}(q))}
   \nonumber \\
   &&+\pi\varepsilon(q^2-\mbox{Re}\Pi^T(q))\mbox{Im}\Pi^T(q)
   \delta((q^2-\Pi^T(q))(q^2-{\Pi^T}^{*}(q))),\\
   \mbox{Im} D^L(q)&=& \mathcal{P}
   \frac{q^2-\mbox{Re}\Pi^L(q)}{(q^2-\Pi^L(q))(q^2-{\Pi^L}^{*}(q))}
   \nonumber \\
   &&+\pi\varepsilon(q^2-\mbox{Re}\Pi^L(q))\mbox{Im}\Pi^L(q)
   \delta((q^2-\Pi^L(q))(q^2-{\Pi^L}^{*}(q))).
\end{eqnarray}
Finally the Schwinger-Dyson equation at finite temperature in the real 
time formalism takes the following form
\begin{eqnarray}
   M(p_0,|\vec{p}|)&=&\frac{1}{\mbox{tr}1}\mbox{tr} \mbox{Re}
   \left[-ie^2\int\frac{d^4q}{(2\pi)^4}
   \gamma^{\mu}iS^{11}(q)\gamma^{\nu}iD^{11}_{\mu\nu}(p-q).\right]
   \nonumber \\
   &=& -e^2\int\frac{d^4q}{(2\pi)^4}M(q_0,|\vec{q}|)\nonumber \\
   && \times
       \left[\mathcal{P}\frac{1}{q^2-M^2(q_0,|\vec{q}|)}
       (2\mbox{Re}D^T(p-q)+\mbox{Re}D^L(p-q))
       \coth\frac{\beta|p_0-q_0|}{2}\right.\nonumber \\
   &&  \left.+(2\mbox{Im}D^T(p-q)+\mbox{Im}D^L(p-q))
       \delta(q^2-M^2(q_0,|\vec{q}|))\tanh\frac{\beta|q_0|}{2}
       \right] ,
\end{eqnarray}
where it should be noted that the following formulae have been employed,
\begin{equation}
   \gamma^{\mu}\gamma^{\nu}P_{\mu\nu}^{T}=-2 , \ \ \ \ \
   \gamma^{\mu}\gamma^{\nu}P_{\mu\nu}^{L}=-1 .
\end{equation}

\section{Instantaneous exchange approximation}

In order to derive informations on the phase structure of quantum
electrodynamics at finite temperature as much as possible we try to
solve the Schwinger-Dyson equation regarding it as an integral
equation for the mass function $M(p_0,|\vec{p}|)$. To solve the
equation it is inevitable to make some approximation which may not
give any serious influence on the resulting physical predictions.
Throughout the paper we apply the instantaneous exchange approximation
in which the $p_0$ dependence of the relevant Green functions
is assumed to be weak and is neglected.
For photons the approximation implies that the vacuum polarization
function is $p_0$ independent. As a functional form of the vacuum
polarization function we adopt the one suggested by the hard thermal
1-loop calculation:
\begin{equation}
\Pi^T(q_0,q)|_{q_0\rightarrow 0}\sim 0, \ \ \ \ \
\Pi^L(q_0,q)|_{q_0\rightarrow 0}
\sim 2Nm_{ph}^2\equiv \frac{N}{3}e^2T^2, \label{VP}
\end{equation}
where $N$ is the number of fermion flavors.
With this approximation the Schwinger-Dyson equation is rewritten as
\begin{eqnarray}
   M(p_0,|\vec{p}|)&=& \pi e^2 \int\frac{d^4 q}{(2\pi)^4}M(q_0,|\vec{q}|)
                    \left(2\mathcal{P}\frac{1}{(\vec{p}-\vec{q})^2}
                    +\mathcal{P}\frac{1}{(\vec{p}-\vec{q})^2+2N{m_{ph}}^2}\
                    \right) \nonumber \\
   & & \times \delta(q^2-M(q_0,|\vec{q}|)^2)\tanh\frac{\beta|q_0|}{2} .
\label{SD:IE}
\end{eqnarray}
Since the right-hand side of Eq.(\ref{SD:IE}) has no $p_0$ dependence
in this approximation, the mass function $M(p_0,|\vec{p}|)$ is
independent of $p_0$ and so is written as $M(|\vec{p}|)$. Hence we obtain
\begin{eqnarray}
   M(p)&=& \frac{\alpha}{2\pi}\int q^2 dq
                   \frac{M(q)}{\sqrt{q^2+M(q)^2}}
   \nonumber \\
   & & \times \frac{1}{2pq}
       \left(2\ln\frac{(p+q)^2}{(p-q)^2}
             +\ln\frac{(p+q)^2+2N{m_{ph}}^2}{(p-q)^2
             +2N{m_{ph}}^2}\right)
              \tanh\frac{\beta\sqrt{q^2+M(q)^2}}{2} \, ,
\label{SD:IE:fin}
\end{eqnarray}
where we set $p=|\vec{p}|$ and $q=|\vec{q}|$.
In the following sections we analyze Eq. (\ref{SD:IE:fin}) numerically 
by using computers and study the behavior of the mass functions to derive 
informations on the phase transitions.

\section{Numerical solutions}

     We would like to solve Eq. (\ref{SD:IE:fin}) by the use of the 
numerical method. There are several different methods available for
solving the integral equation (\ref{SD:IE:fin}) numerically.
Among those methods there exist two standard methods. The one is to
discretize the integral in Eq. (\ref{SD:IE:fin}) and regard the
resulting equation as simultaneous equations for the mass function.
The other of the two is to start with the suitable trial
function for the solution and to iterate it until we reach a stable
solution. It seems that the latter method is much easier to handle
and is useful as long as the convergence of the iteration is
guaranteed. We would like to adopt this latter method in the following
arguments.

     We begin with the simplest possible choice for the trial mass
function in the iteration. Thus the trial function is chosen to be
a constant independent of $p$,
\begin{equation}
M(p)=\textrm{constant}.
\end{equation}
At each step of the iteration the integration is performed by
the use of the Monte Carlo method and the integral is cut off
at mass scale $\Lambda$.
After the first iteration the resulting mass function acquires a
$p$ dependence and is substituted for the mass function in the
integral on the right hand side of Eq. (\ref{SD:IE:fin}). Repeating
this procedure we may have a possible stable result. Whether we
have a stable result or not should be always checked after
sufficiently many iterations. In each calculation in the following
we confirm the stability of the solution, that is, we obtain the
same result starting from the different trial functions.

     We first consider the case without the thermal photon mass
$N=0$. Note here that parameter $N$ plays a role of switching on
($N=1$) and off ($N=0$) the photon mass and also represents the
number of fermion flavors.
In Fig. 1 the mass function normalized by the cut-off $\Lambda$
of the $p$ integration is presented as a function of the number
of the iterations for the case with $p=\Lambda/100$
and $T=0.01\Lambda, 0.20\Lambda, 0.25\Lambda$. Here we adopt
the unit system with $k=1$ where $k$ is the Boltzmann constant.
The fluctuations observed in Fig. 1 are given rise to by the errors
in the Monte Carlo integration.

\begin{figure}
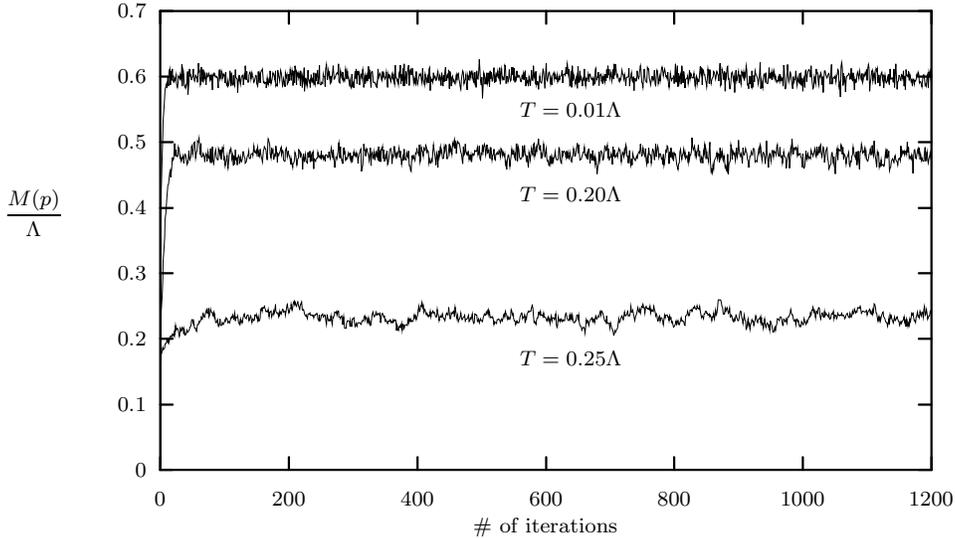

\begin{center}
\setlength{\unitlength}{0.240900pt}


\caption{Typical behavior of the solution of Schwinger-Dyson equation
(\ref{SD:IE:fin}) for $N=0$, $\alpha=1$, and $p=0.1\Lambda$.}
\end{center}
\end{figure}

As is seen in Fig. 1, the resulting mass function becomes stable
after about 100 iterations. We then push forward our analysis
by taking more points for the values of the momentum and
obtain the momentum dependence of the solution for the mass
function. In Fig. 2 the mass function normalized by the cut-off
is presented as a function of the momentum $p$ normalized by
the cut-off. The mass functions were obtained after 1200 iterations.
It should be noted here that the convergence of the iterations
becomes slower if we sit near the critical value of the coupling
constant $\alpha$ and temperature $T$. In this case we need more
iterations to obtain a stable result for the mass function.

\begin{figure}
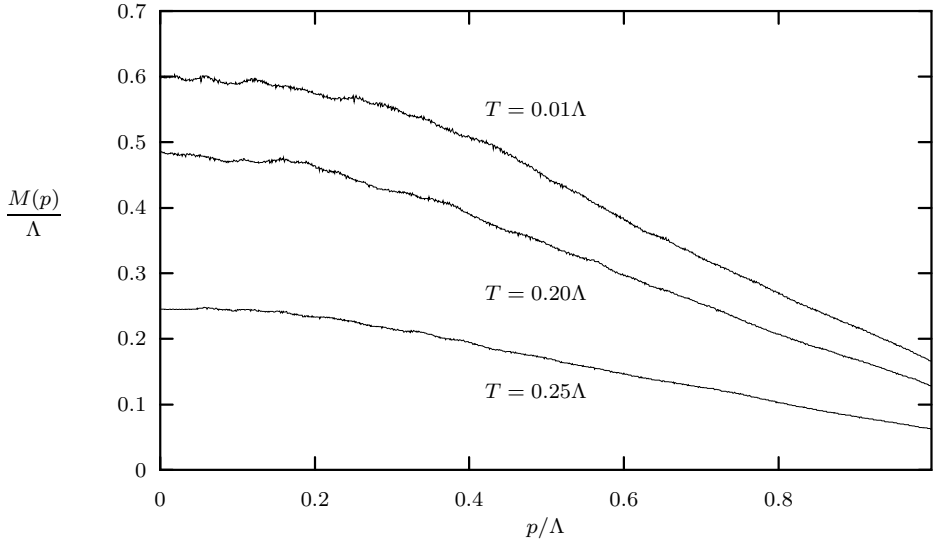

\begin{center}
\setlength{\unitlength}{0.240900pt}


\caption{Typical shape of the mass function $M(p)$ for $N=0$ 
and $\alpha=1$.}
\end{center}
\end{figure}

The case with the thermal photon mass $N=1$ is studied essentially
in the same way as above. We may also study the case with three
fermion flavors $N=3$ essentially in the same manner. Here we skip
to present the numerical results on the mass function.

\section{Chiral phase transitions}

     We first wish to observe the behavior of the mass function at
some fixed value of $p$ as a function of parameters $\alpha$ and $T$.
The $\alpha$ dependence of the mass function $M(p)$ with $p=0.1\Lambda$
is shown in Fig. 3 for various fixed values of $T$. Note here that the
errors coming from the fluctuations observed in Fig. 1 are smaller than
the size of the mark shown for each sample point in Fig. 3.

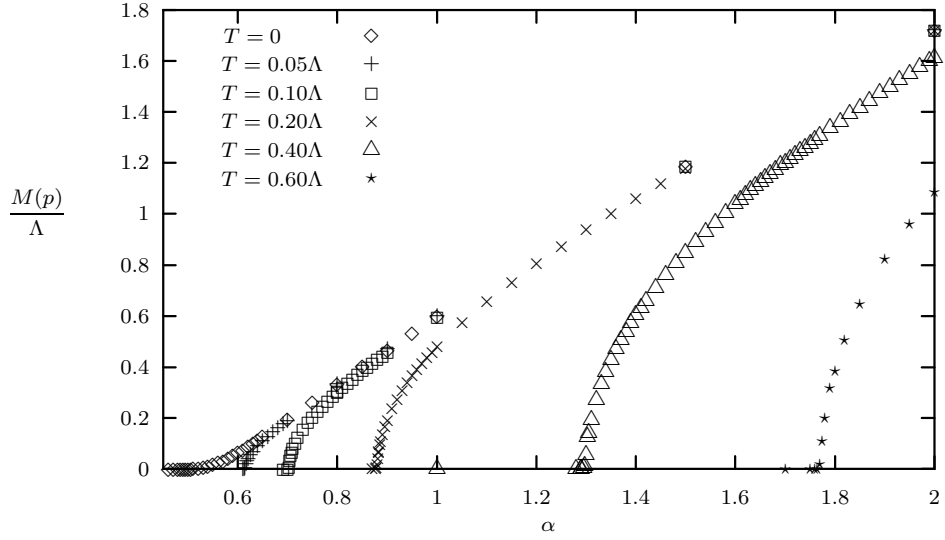
\begin{figure}
\begin{center}
\setlength{\unitlength}{0.240900pt}
\begin{picture}(1500,900)(0,0)
\footnotesize
\thicklines \path(245,135)(265,135)
\thicklines \path(1456,135)(1436,135)
\put(223,135){\makebox(0,0)[r]{0}}
\thicklines \path(245,215)(265,215)
\thicklines \path(1456,215)(1436,215)
\put(223,215){\makebox(0,0)[r]{0.2}}
\thicklines \path(245,295)(265,295)
\thicklines \path(1456,295)(1436,295)
\put(223,295){\makebox(0,0)[r]{0.4}}
\thicklines \path(245,375)(265,375)
\thicklines \path(1456,375)(1436,375)
\put(223,375){\makebox(0,0)[r]{0.6}}
\thicklines \path(245,455)(265,455)
\thicklines \path(1456,455)(1436,455)
\put(223,455){\makebox(0,0)[r]{0.8}}
\thicklines \path(245,536)(265,536)
\thicklines \path(1456,536)(1436,536)
\put(223,536){\makebox(0,0)[r]{1}}
\thicklines \path(245,616)(265,616)
\thicklines \path(1456,616)(1436,616)
\put(223,616){\makebox(0,0)[r]{1.2}}
\thicklines \path(245,696)(265,696)
\thicklines \path(1456,696)(1436,696)
\put(223,696){\makebox(0,0)[r]{1.4}}
\thicklines \path(245,776)(265,776)
\thicklines \path(1456,776)(1436,776)
\put(223,776){\makebox(0,0)[r]{1.6}}
\thicklines \path(245,856)(265,856)
\thicklines \path(1456,856)(1436,856)
\put(223,856){\makebox(0,0)[r]{1.8}}
\thicklines \path(362,135)(362,155)
\thicklines \path(362,856)(362,836)
\put(362,90){\makebox(0,0){0.6}}
\thicklines \path(518,135)(518,155)
\thicklines \path(518,856)(518,836)
\put(518,90){\makebox(0,0){0.8}}
\thicklines \path(675,135)(675,155)
\thicklines \path(675,856)(675,836)
\put(675,90){\makebox(0,0){1}}
\thicklines \path(831,135)(831,155)
\thicklines \path(831,856)(831,836)
\put(831,90){\makebox(0,0){1.2}}
\thicklines \path(987,135)(987,155)
\thicklines \path(987,856)(987,836)
\put(987,90){\makebox(0,0){1.4}}
\thicklines \path(1143,135)(1143,155)
\thicklines \path(1143,856)(1143,836)
\put(1143,90){\makebox(0,0){1.6}}
\thicklines \path(1300,135)(1300,155)
\thicklines \path(1300,856)(1300,836)
\put(1300,90){\makebox(0,0){1.8}}
\thicklines \path(1456,135)(1456,155)
\thicklines \path(1456,856)(1456,836)
\put(1456,90){\makebox(0,0){2}}
\thicklines \path(245,135)(1456,135)(1456,856)(245,856)(245,135)
\put(0,539){\makebox(0,0)[l]
{\shortstack{$\displaystyle\frac{M(p)}{\Lambda}$}}}
\put(850,45){\makebox(0,0){$\alpha$}}
\put(496,816){\makebox(0,0)[r]{$T=0\hspace{4.6ex}$}}
\put(253,135){\raisebox{-1.2pt}{\makebox(0,0){$\Diamond$}}}
\put(261,135){\raisebox{-1.2pt}{\makebox(0,0){$\Diamond$}}}
\put(268,135){\raisebox{-1.2pt}{\makebox(0,0){$\Diamond$}}}
\put(272,135){\raisebox{-1.2pt}{\makebox(0,0){$\Diamond$}}}
\put(276,135){\raisebox{-1.2pt}{\makebox(0,0){$\Diamond$}}}
\put(280,135){\raisebox{-1.2pt}{\makebox(0,0){$\Diamond$}}}
\put(284,135){\raisebox{-1.2pt}{\makebox(0,0){$\Diamond$}}}
\put(288,135){\raisebox{-1.2pt}{\makebox(0,0){$\Diamond$}}}
\put(292,136){\raisebox{-1.2pt}{\makebox(0,0){$\Diamond$}}}
\put(300,136){\raisebox{-1.2pt}{\makebox(0,0){$\Diamond$}}}
\put(308,138){\raisebox{-1.2pt}{\makebox(0,0){$\Diamond$}}}
\put(315,139){\raisebox{-1.2pt}{\makebox(0,0){$\Diamond$}}}
\put(323,142){\raisebox{-1.2pt}{\makebox(0,0){$\Diamond$}}}
\put(331,145){\raisebox{-1.2pt}{\makebox(0,0){$\Diamond$}}}
\put(339,148){\raisebox{-1.2pt}{\makebox(0,0){$\Diamond$}}}
\put(347,152){\raisebox{-1.2pt}{\makebox(0,0){$\Diamond$}}}
\put(354,157){\raisebox{-1.2pt}{\makebox(0,0){$\Diamond$}}}
\put(362,162){\raisebox{-1.2pt}{\makebox(0,0){$\Diamond$}}}
\put(370,165){\raisebox{-1.2pt}{\makebox(0,0){$\Diamond$}}}
\put(378,171){\raisebox{-1.2pt}{\makebox(0,0){$\Diamond$}}}
\put(386,176){\raisebox{-1.2pt}{\makebox(0,0){$\Diamond$}}}
\put(393,181){\raisebox{-1.2pt}{\makebox(0,0){$\Diamond$}}}
\put(401,186){\raisebox{-1.2pt}{\makebox(0,0){$\Diamond$}}}
\put(440,213){\raisebox{-1.2pt}{\makebox(0,0){$\Diamond$}}}
\put(479,240){\raisebox{-1.2pt}{\makebox(0,0){$\Diamond$}}}
\put(518,268){\raisebox{-1.2pt}{\makebox(0,0){$\Diamond$}}}
\put(558,296){\raisebox{-1.2pt}{\makebox(0,0){$\Diamond$}}}
\put(597,322){\raisebox{-1.2pt}{\makebox(0,0){$\Diamond$}}}
\put(636,349){\raisebox{-1.2pt}{\makebox(0,0){$\Diamond$}}}
\put(675,375){\raisebox{-1.2pt}{\makebox(0,0){$\Diamond$}}}
\put(1065,611){\raisebox{-1.2pt}{\makebox(0,0){$\Diamond$}}}
\put(1456,825){\raisebox{-1.2pt}{\makebox(0,0){$\Diamond$}}}
\put(572,816){\raisebox{-1.2pt}{\makebox(0,0){$\Diamond$}}}
\put(496,771){\makebox(0,0)[r]{$T=0.05\Lambda$}}
\put(370,135){\makebox(0,0){$+$}}
\put(372,135){\makebox(0,0){$+$}}
\put(373,136){\makebox(0,0){$+$}}
\put(375,142){\makebox(0,0){$+$}}
\put(376,148){\makebox(0,0){$+$}}
\put(378,150){\makebox(0,0){$+$}}
\put(382,157){\makebox(0,0){$+$}}
\put(386,162){\makebox(0,0){$+$}}
\put(393,171){\makebox(0,0){$+$}}
\put(401,179){\makebox(0,0){$+$}}
\put(409,186){\makebox(0,0){$+$}}
\put(417,193){\makebox(0,0){$+$}}
\put(425,199){\makebox(0,0){$+$}}
\put(433,205){\makebox(0,0){$+$}}
\put(440,211){\makebox(0,0){$+$}}
\put(518,268){\makebox(0,0){$+$}}
\put(597,323){\makebox(0,0){$+$}}
\put(675,375){\makebox(0,0){$+$}}
\put(1456,825){\makebox(0,0){$+$}}
\put(572,771){\makebox(0,0){$+$}}
\put(496,726){\makebox(0,0)[r]{$T=0.10\Lambda$}}
\put(433,135){\raisebox{-1.2pt}{\makebox(0,0){$\Box$}}}
\put(440,135){\raisebox{-1.2pt}{\makebox(0,0){$\Box$}}}
\put(442,138){\raisebox{-1.2pt}{\makebox(0,0){$\Box$}}}
\put(443,149){\raisebox{-1.2pt}{\makebox(0,0){$\Box$}}}
\put(445,156){\raisebox{-1.2pt}{\makebox(0,0){$\Box$}}}
\put(447,162){\raisebox{-1.2pt}{\makebox(0,0){$\Box$}}}
\put(448,167){\raisebox{-1.2pt}{\makebox(0,0){$\Box$}}}
\put(452,176){\raisebox{-1.2pt}{\makebox(0,0){$\Box$}}}
\put(456,185){\raisebox{-1.2pt}{\makebox(0,0){$\Box$}}}
\put(464,197){\raisebox{-1.2pt}{\makebox(0,0){$\Box$}}}
\put(472,208){\raisebox{-1.2pt}{\makebox(0,0){$\Box$}}}
\put(479,217){\raisebox{-1.2pt}{\makebox(0,0){$\Box$}}}
\put(487,226){\raisebox{-1.2pt}{\makebox(0,0){$\Box$}}}
\put(495,234){\raisebox{-1.2pt}{\makebox(0,0){$\Box$}}}
\put(503,242){\raisebox{-1.2pt}{\makebox(0,0){$\Box$}}}
\put(511,249){\raisebox{-1.2pt}{\makebox(0,0){$\Box$}}}
\put(518,256){\raisebox{-1.2pt}{\makebox(0,0){$\Box$}}}
\put(518,257){\raisebox{-1.2pt}{\makebox(0,0){$\Box$}}}
\put(526,264){\raisebox{-1.2pt}{\makebox(0,0){$\Box$}}}
\put(534,270){\raisebox{-1.2pt}{\makebox(0,0){$\Box$}}}
\put(542,276){\raisebox{-1.2pt}{\makebox(0,0){$\Box$}}}
\put(550,284){\raisebox{-1.2pt}{\makebox(0,0){$\Box$}}}
\put(558,290){\raisebox{-1.2pt}{\makebox(0,0){$\Box$}}}
\put(565,295){\raisebox{-1.2pt}{\makebox(0,0){$\Box$}}}
\put(573,302){\raisebox{-1.2pt}{\makebox(0,0){$\Box$}}}
\put(581,307){\raisebox{-1.2pt}{\makebox(0,0){$\Box$}}}
\put(589,313){\raisebox{-1.2pt}{\makebox(0,0){$\Box$}}}
\put(597,319){\raisebox{-1.2pt}{\makebox(0,0){$\Box$}}}
\put(675,373){\raisebox{-1.2pt}{\makebox(0,0){$\Box$}}}
\put(1065,611){\raisebox{-1.2pt}{\makebox(0,0){$\Box$}}}
\put(1456,825){\raisebox{-1.2pt}{\makebox(0,0){$\Box$}}}
\put(572,726){\raisebox{-1.2pt}{\makebox(0,0){$\Box$}}}
\put(496,681){\makebox(0,0)[r]{$T=0.20\Lambda$}}
\put(573,135){\makebox(0,0){$\times$}}
\put(578,135){\makebox(0,0){$\times$}}
\put(579,137){\makebox(0,0){$\times$}}
\put(581,141){\makebox(0,0){$\times$}}
\put(582,151){\makebox(0,0){$\times$}}
\put(583,162){\makebox(0,0){$\times$}}
\put(583,160){\makebox(0,0){$\times$}}
\put(585,173){\makebox(0,0){$\times$}}
\put(586,178){\makebox(0,0){$\times$}}
\put(589,188){\makebox(0,0){$\times$}}
\put(593,201){\makebox(0,0){$\times$}}
\put(597,211){\makebox(0,0){$\times$}}
\put(604,229){\makebox(0,0){$\times$}}
\put(612,244){\makebox(0,0){$\times$}}
\put(620,258){\makebox(0,0){$\times$}}
\put(628,270){\makebox(0,0){$\times$}}
\put(636,281){\makebox(0,0){$\times$}}
\put(643,291){\makebox(0,0){$\times$}}
\put(651,300){\makebox(0,0){$\times$}}
\put(659,309){\makebox(0,0){$\times$}}
\put(667,318){\makebox(0,0){$\times$}}
\put(675,327){\makebox(0,0){$\times$}}
\put(714,365){\makebox(0,0){$\times$}}
\put(753,398){\makebox(0,0){$\times$}}
\put(792,428){\makebox(0,0){$\times$}}
\put(831,457){\makebox(0,0){$\times$}}
\put(870,484){\makebox(0,0){$\times$}}
\put(909,510){\makebox(0,0){$\times$}}
\put(948,535){\makebox(0,0){$\times$}}
\put(987,559){\makebox(0,0){$\times$}}
\put(1026,583){\makebox(0,0){$\times$}}
\put(1065,606){\makebox(0,0){$\times$}}
\put(1456,824){\makebox(0,0){$\times$}}
\put(572,681){\makebox(0,0){$\times$}}
\put(496,636){\makebox(0,0)[r]{$T=0.40\Lambda$}}
\put(675,135){\makebox(0,0){$\triangle$}}
\put(893,135){\makebox(0,0){$\triangle$}}
\put(901,135){\makebox(0,0){$\triangle$}}
\put(901,135){\makebox(0,0){$\triangle$}}
\put(905,136){\makebox(0,0){$\triangle$}}
\put(907,140){\makebox(0,0){$\triangle$}}
\put(909,157){\makebox(0,0){$\triangle$}}
\put(911,185){\makebox(0,0){$\triangle$}}
\put(913,191){\makebox(0,0){$\triangle$}}
\put(917,212){\makebox(0,0){$\triangle$}}
\put(925,243){\makebox(0,0){$\triangle$}}
\put(933,268){\makebox(0,0){$\triangle$}}
\put(940,288){\makebox(0,0){$\triangle$}}
\put(948,306){\makebox(0,0){$\triangle$}}
\put(956,323){\makebox(0,0){$\triangle$}}
\put(964,338){\makebox(0,0){$\triangle$}}
\put(972,351){\makebox(0,0){$\triangle$}}
\put(979,364){\makebox(0,0){$\triangle$}}
\put(987,377){\makebox(0,0){$\triangle$}}
\put(995,388){\makebox(0,0){$\triangle$}}
\put(1003,399){\makebox(0,0){$\triangle$}}
\put(1018,420){\makebox(0,0){$\triangle$}}
\put(1034,440){\makebox(0,0){$\triangle$}}
\put(1050,458){\makebox(0,0){$\triangle$}}
\put(1065,475){\makebox(0,0){$\triangle$}}
\put(1081,492){\makebox(0,0){$\triangle$}}
\put(1097,507){\makebox(0,0){$\triangle$}}
\put(1112,522){\makebox(0,0){$\triangle$}}
\put(1128,537){\makebox(0,0){$\triangle$}}
\put(1143,551){\makebox(0,0){$\triangle$}}
\put(1151,558){\makebox(0,0){$\triangle$}}
\put(1159,566){\makebox(0,0){$\triangle$}}
\put(1167,573){\makebox(0,0){$\triangle$}}
\put(1175,580){\makebox(0,0){$\triangle$}}
\put(1183,586){\makebox(0,0){$\triangle$}}
\put(1190,592){\makebox(0,0){$\triangle$}}
\put(1198,599){\makebox(0,0){$\triangle$}}
\put(1206,605){\makebox(0,0){$\triangle$}}
\put(1214,612){\makebox(0,0){$\triangle$}}
\put(1222,616){\makebox(0,0){$\triangle$}}
\put(1229,622){\makebox(0,0){$\triangle$}}
\put(1237,628){\makebox(0,0){$\triangle$}}
\put(1245,634){\makebox(0,0){$\triangle$}}
\put(1253,640){\makebox(0,0){$\triangle$}}
\put(1261,646){\makebox(0,0){$\triangle$}}
\put(1268,652){\makebox(0,0){$\triangle$}}
\put(1276,659){\makebox(0,0){$\triangle$}}
\put(1292,671){\makebox(0,0){$\triangle$}}
\put(1308,681){\makebox(0,0){$\triangle$}}
\put(1323,693){\makebox(0,0){$\triangle$}}
\put(1339,703){\makebox(0,0){$\triangle$}}
\put(1354,714){\makebox(0,0){$\triangle$}}
\put(1370,726){\makebox(0,0){$\triangle$}}
\put(1386,736){\makebox(0,0){$\triangle$}}
\put(1401,746){\makebox(0,0){$\triangle$}}
\put(1417,756){\makebox(0,0){$\triangle$}}
\put(1433,766){\makebox(0,0){$\triangle$}}
\put(1448,776){\makebox(0,0){$\triangle$}}
\put(1456,781){\makebox(0,0){$\triangle$}}
\put(572,636){\makebox(0,0){$\triangle$}}
\put(496,591){\makebox(0,0)[r]{$T=0.60\Lambda$}}
\put(1222,135){\makebox(0,0){$\star$}}
\put(1261,135){\makebox(0,0){$\star$}}
\put(1268,135){\makebox(0,0){$\star$}}
\put(1272,135){\makebox(0,0){$\star$}}
\put(1276,142){\makebox(0,0){$\star$}}
\put(1280,178){\makebox(0,0){$\star$}}
\put(1284,214){\makebox(0,0){$\star$}}
\put(1292,262){\makebox(0,0){$\star$}}
\put(1300,289){\makebox(0,0){$\star$}}
\put(1315,337){\makebox(0,0){$\star$}}
\put(1339,394){\makebox(0,0){$\star$}}
\put(1378,464){\makebox(0,0){$\star$}}
\put(1417,520){\makebox(0,0){$\star$}}
\put(1456,569){\makebox(0,0){$\star$}}
\put(572,591){\makebox(0,0){$\star$}}
\end{picture}
\caption{Dynamical fermion mass as a function of the coupling constant 
$\alpha$ for $N=0$ at $p=0.1\Lambda$}
\end{center}
\end{figure}
\begin{figure}
\begin{center}
\setlength{\unitlength}{0.240900pt}
\begin{picture}(1500,900)(0,0)
\footnotesize
\thicklines \path(245,135)(265,135)
\thicklines \path(1456,135)(1436,135)
\put(223,135){\makebox(0,0)[r]{0}}
\thicklines \path(245,225)(265,225)
\thicklines \path(1456,225)(1436,225)
\put(223,225){\makebox(0,0)[r]{0.5}}
\thicklines \path(245,315)(265,315)
\thicklines \path(1456,315)(1436,315)
\put(223,315){\makebox(0,0)[r]{1}}
\thicklines \path(245,405)(265,405)
\thicklines \path(1456,405)(1436,405)
\put(223,405){\makebox(0,0)[r]{1.5}}
\thicklines \path(245,496)(265,496)
\thicklines \path(1456,496)(1436,496)
\put(223,496){\makebox(0,0)[r]{2}}
\thicklines \path(245,586)(265,586)
\thicklines \path(1456,586)(1436,586)
\put(223,586){\makebox(0,0)[r]{2.5}}
\thicklines \path(245,676)(265,676)
\thicklines \path(1456,676)(1436,676)
\put(223,676){\makebox(0,0)[r]{3}}
\thicklines \path(245,766)(265,766)
\thicklines \path(1456,766)(1436,766)
\put(223,766){\makebox(0,0)[r]{3.5}}
\thicklines \path(245,856)(265,856)
\thicklines \path(1456,856)(1436,856)
\put(223,856){\makebox(0,0)[r]{4}}
\thicklines \path(245,135)(245,155)
\thicklines \path(245,856)(245,836)
\put(245,90){\makebox(0,0){0}}
\thicklines \path(487,135)(487,155)
\thicklines \path(487,856)(487,836)
\put(487,90){\makebox(0,0){0.2}}
\thicklines \path(729,135)(729,155)
\thicklines \path(729,856)(729,836)
\put(729,90){\makebox(0,0){0.4}}
\thicklines \path(972,135)(972,155)
\thicklines \path(972,856)(972,836)
\put(972,90){\makebox(0,0){0.6}}
\thicklines \path(1214,135)(1214,155)
\thicklines \path(1214,856)(1214,836)
\put(1214,90){\makebox(0,0){0.8}}
\thicklines \path(1456,135)(1456,155)
\thicklines \path(1456,856)(1456,836)
\put(1456,90){\makebox(0,0){1}}
\thicklines \path(245,135)(1456,135)(1456,856)(245,856)(245,135)
\put(0,539){\makebox(0,0)[l]
{\shortstack{$\displaystyle\frac{M(p)}{\Lambda}$}}}
\put(850,45){\makebox(0,0){$T/\Lambda$}}
\put(1313,351){\makebox(0,0)[r]{$\alpha=0.8$}}
\put(245,195){\raisebox{-1.2pt}{\makebox(0,0){$\Diamond$}}}
\put(251,195){\raisebox{-1.2pt}{\makebox(0,0){$\Diamond$}}}
\put(257,195){\raisebox{-1.2pt}{\makebox(0,0){$\Diamond$}}}
\put(263,195){\raisebox{-1.2pt}{\makebox(0,0){$\Diamond$}}}
\put(269,195){\raisebox{-1.2pt}{\makebox(0,0){$\Diamond$}}}
\put(281,195){\raisebox{-1.2pt}{\makebox(0,0){$\Diamond$}}}
\put(291,195){\raisebox{-1.2pt}{\makebox(0,0){$\Diamond$}}}
\put(293,195){\raisebox{-1.2pt}{\makebox(0,0){$\Diamond$}}}
\put(306,195){\raisebox{-1.2pt}{\makebox(0,0){$\Diamond$}}}
\put(318,195){\raisebox{-1.2pt}{\makebox(0,0){$\Diamond$}}}
\put(330,194){\raisebox{-1.2pt}{\makebox(0,0){$\Diamond$}}}
\put(342,193){\raisebox{-1.2pt}{\makebox(0,0){$\Diamond$}}}
\put(348,193){\raisebox{-1.2pt}{\makebox(0,0){$\Diamond$}}}
\put(354,192){\raisebox{-1.2pt}{\makebox(0,0){$\Diamond$}}}
\put(360,191){\raisebox{-1.2pt}{\makebox(0,0){$\Diamond$}}}
\put(366,190){\raisebox{-1.2pt}{\makebox(0,0){$\Diamond$}}}
\put(366,190){\raisebox{-1.2pt}{\makebox(0,0){$\Diamond$}}}
\put(372,189){\raisebox{-1.2pt}{\makebox(0,0){$\Diamond$}}}
\put(378,187){\raisebox{-1.2pt}{\makebox(0,0){$\Diamond$}}}
\put(384,185){\raisebox{-1.2pt}{\makebox(0,0){$\Diamond$}}}
\put(390,183){\raisebox{-1.2pt}{\makebox(0,0){$\Diamond$}}}
\put(402,177){\raisebox{-1.2pt}{\makebox(0,0){$\Diamond$}}}
\put(408,174){\raisebox{-1.2pt}{\makebox(0,0){$\Diamond$}}}
\put(415,170){\raisebox{-1.2pt}{\makebox(0,0){$\Diamond$}}}
\put(421,164){\raisebox{-1.2pt}{\makebox(0,0){$\Diamond$}}}
\put(427,158){\raisebox{-1.2pt}{\makebox(0,0){$\Diamond$}}}
\put(429,154){\raisebox{-1.2pt}{\makebox(0,0){$\Diamond$}}}
\put(433,145){\raisebox{-1.2pt}{\makebox(0,0){$\Diamond$}}}
\put(435,136){\raisebox{-1.2pt}{\makebox(0,0){$\Diamond$}}}
\put(439,135){\raisebox{-1.2pt}{\makebox(0,0){$\Diamond$}}}
\put(445,135){\raisebox{-1.2pt}{\makebox(0,0){$\Diamond$}}}
\put(451,135){\raisebox{-1.2pt}{\makebox(0,0){$\Diamond$}}}
\put(1389,351){\raisebox{-1.2pt}{\makebox(0,0){$\Diamond$}}}
\put(1313,306){\makebox(0,0)[r]{$\alpha=1.0$}}
\put(245,243){\makebox(0,0){$+$}}
\put(251,243){\makebox(0,0){$+$}}
\put(257,243){\makebox(0,0){$+$}}
\put(263,243){\makebox(0,0){$+$}}
\put(269,243){\makebox(0,0){$+$}}
\put(281,243){\makebox(0,0){$+$}}
\put(306,243){\makebox(0,0){$+$}}
\put(330,243){\makebox(0,0){$+$}}
\put(354,243){\makebox(0,0){$+$}}
\put(366,242){\makebox(0,0){$+$}}
\put(390,241){\makebox(0,0){$+$}}
\put(402,240){\makebox(0,0){$+$}}
\put(415,239){\makebox(0,0){$+$}}
\put(427,237){\makebox(0,0){$+$}}
\put(433,236){\makebox(0,0){$+$}}
\put(439,235){\makebox(0,0){$+$}}
\put(445,234){\makebox(0,0){$+$}}
\put(451,233){\makebox(0,0){$+$}}
\put(457,231){\makebox(0,0){$+$}}
\put(463,229){\makebox(0,0){$+$}}
\put(475,226){\makebox(0,0){$+$}}
\put(487,221){\makebox(0,0){$+$}}
\put(499,216){\makebox(0,0){$+$}}
\put(511,210){\makebox(0,0){$+$}}
\put(517,205){\makebox(0,0){$+$}}
\put(524,202){\makebox(0,0){$+$}}
\put(536,191){\makebox(0,0){$+$}}
\put(548,177){\makebox(0,0){$+$}}
\put(554,167){\makebox(0,0){$+$}}
\put(556,162){\makebox(0,0){$+$}}
\put(560,150){\makebox(0,0){$+$}}
\put(562,137){\makebox(0,0){$+$}}
\put(566,135){\makebox(0,0){$+$}}
\put(578,135){\makebox(0,0){$+$}}
\put(608,135){\makebox(0,0){$+$}}
\put(639,135){\makebox(0,0){$+$}}
\put(669,135){\makebox(0,0){$+$}}
\put(699,135){\makebox(0,0){$+$}}
\put(1389,306){\makebox(0,0){$+$}}
\put(1313,261){\makebox(0,0)[r]{$\alpha=2.0$}}
\put(245,445){\raisebox{-1.2pt}{\makebox(0,0){$\Box$}}}
\put(251,446){\raisebox{-1.2pt}{\makebox(0,0){$\Box$}}}
\put(257,446){\raisebox{-1.2pt}{\makebox(0,0){$\Box$}}}
\put(306,445){\raisebox{-1.2pt}{\makebox(0,0){$\Box$}}}
\put(366,445){\raisebox{-1.2pt}{\makebox(0,0){$\Box$}}}
\put(427,445){\raisebox{-1.2pt}{\makebox(0,0){$\Box$}}}
\put(487,445){\raisebox{-1.2pt}{\makebox(0,0){$\Box$}}}
\put(548,444){\raisebox{-1.2pt}{\makebox(0,0){$\Box$}}}
\put(608,441){\raisebox{-1.2pt}{\makebox(0,0){$\Box$}}}
\put(669,435){\raisebox{-1.2pt}{\makebox(0,0){$\Box$}}}
\put(729,426){\raisebox{-1.2pt}{\makebox(0,0){$\Box$}}}
\put(742,423){\raisebox{-1.2pt}{\makebox(0,0){$\Box$}}}
\put(754,421){\raisebox{-1.2pt}{\makebox(0,0){$\Box$}}}
\put(766,419){\raisebox{-1.2pt}{\makebox(0,0){$\Box$}}}
\put(778,415){\raisebox{-1.2pt}{\makebox(0,0){$\Box$}}}
\put(790,413){\raisebox{-1.2pt}{\makebox(0,0){$\Box$}}}
\put(802,409){\raisebox{-1.2pt}{\makebox(0,0){$\Box$}}}
\put(814,406){\raisebox{-1.2pt}{\makebox(0,0){$\Box$}}}
\put(826,402){\raisebox{-1.2pt}{\makebox(0,0){$\Box$}}}
\put(838,398){\raisebox{-1.2pt}{\makebox(0,0){$\Box$}}}
\put(851,394){\raisebox{-1.2pt}{\makebox(0,0){$\Box$}}}
\put(863,390){\raisebox{-1.2pt}{\makebox(0,0){$\Box$}}}
\put(875,385){\raisebox{-1.2pt}{\makebox(0,0){$\Box$}}}
\put(887,379){\raisebox{-1.2pt}{\makebox(0,0){$\Box$}}}
\put(899,374){\raisebox{-1.2pt}{\makebox(0,0){$\Box$}}}
\put(911,367){\raisebox{-1.2pt}{\makebox(0,0){$\Box$}}}
\put(923,362){\raisebox{-1.2pt}{\makebox(0,0){$\Box$}}}
\put(935,355){\raisebox{-1.2pt}{\makebox(0,0){$\Box$}}}
\put(947,348){\raisebox{-1.2pt}{\makebox(0,0){$\Box$}}}
\put(959,340){\raisebox{-1.2pt}{\makebox(0,0){$\Box$}}}
\put(972,330){\raisebox{-1.2pt}{\makebox(0,0){$\Box$}}}
\put(984,321){\raisebox{-1.2pt}{\makebox(0,0){$\Box$}}}
\put(996,312){\raisebox{-1.2pt}{\makebox(0,0){$\Box$}}}
\put(1008,301){\raisebox{-1.2pt}{\makebox(0,0){$\Box$}}}
\put(1020,289){\raisebox{-1.2pt}{\makebox(0,0){$\Box$}}}
\put(1032,272){\raisebox{-1.2pt}{\makebox(0,0){$\Box$}}}
\put(1044,259){\raisebox{-1.2pt}{\makebox(0,0){$\Box$}}}
\put(1056,238){\raisebox{-1.2pt}{\makebox(0,0){$\Box$}}}
\put(1068,213){\raisebox{-1.2pt}{\makebox(0,0){$\Box$}}}
\put(1075,188){\raisebox{-1.2pt}{\makebox(0,0){$\Box$}}}
\put(1081,162){\raisebox{-1.2pt}{\makebox(0,0){$\Box$}}}
\put(1082,148){\raisebox{-1.2pt}{\makebox(0,0){$\Box$}}}
\put(1083,145){\raisebox{-1.2pt}{\makebox(0,0){$\Box$}}}
\put(1084,138){\raisebox{-1.2pt}{\makebox(0,0){$\Box$}}}
\put(1085,136){\raisebox{-1.2pt}{\makebox(0,0){$\Box$}}}
\put(1087,135){\raisebox{-1.2pt}{\makebox(0,0){$\Box$}}}
\put(1088,135){\raisebox{-1.2pt}{\makebox(0,0){$\Box$}}}
\put(1090,135){\raisebox{-1.2pt}{\makebox(0,0){$\Box$}}}
\put(1093,135){\raisebox{-1.2pt}{\makebox(0,0){$\Box$}}}
\put(1105,135){\raisebox{-1.2pt}{\makebox(0,0){$\Box$}}}
\put(1117,135){\raisebox{-1.2pt}{\makebox(0,0){$\Box$}}}
\put(1129,135){\raisebox{-1.2pt}{\makebox(0,0){$\Box$}}}
\put(1141,135){\raisebox{-1.2pt}{\makebox(0,0){$\Box$}}}
\put(1389,261){\raisebox{-1.2pt}{\makebox(0,0){$\Box$}}}
\put(1313,216){\makebox(0,0)[r]{$\alpha=3.0$}}
\put(245,628){\makebox(0,0){$\times$}}
\put(251,627){\makebox(0,0){$\times$}}
\put(257,628){\makebox(0,0){$\times$}}
\put(306,627){\makebox(0,0){$\times$}}
\put(366,627){\makebox(0,0){$\times$}}
\put(487,627){\makebox(0,0){$\times$}}
\put(608,627){\makebox(0,0){$\times$}}
\put(729,623){\makebox(0,0){$\times$}}
\put(851,616){\makebox(0,0){$\times$}}
\put(972,600){\makebox(0,0){$\times$}}
\put(1032,588){\makebox(0,0){$\times$}}
\put(1044,586){\makebox(0,0){$\times$}}
\put(1056,583){\makebox(0,0){$\times$}}
\put(1068,581){\makebox(0,0){$\times$}}
\put(1081,578){\makebox(0,0){$\times$}}
\put(1093,574){\makebox(0,0){$\times$}}
\put(1105,571){\makebox(0,0){$\times$}}
\put(1117,568){\makebox(0,0){$\times$}}
\put(1129,563){\makebox(0,0){$\times$}}
\put(1141,560){\makebox(0,0){$\times$}}
\put(1153,557){\makebox(0,0){$\times$}}
\put(1165,553){\makebox(0,0){$\times$}}
\put(1177,549){\makebox(0,0){$\times$}}
\put(1190,544){\makebox(0,0){$\times$}}
\put(1202,540){\makebox(0,0){$\times$}}
\put(1214,536){\makebox(0,0){$\times$}}
\put(1226,531){\makebox(0,0){$\times$}}
\put(1238,526){\makebox(0,0){$\times$}}
\put(1250,520){\makebox(0,0){$\times$}}
\put(1274,508){\makebox(0,0){$\times$}}
\put(1299,497){\makebox(0,0){$\times$}}
\put(1323,484){\makebox(0,0){$\times$}}
\put(1347,469){\makebox(0,0){$\times$}}
\put(1371,453){\makebox(0,0){$\times$}}
\put(1395,433){\makebox(0,0){$\times$}}
\put(1420,412){\makebox(0,0){$\times$}}
\put(1444,392){\makebox(0,0){$\times$}}
\put(1456,375){\makebox(0,0){$\times$}}
\put(1389,216){\makebox(0,0){$\times$}}
\put(1313,171){\makebox(0,0)[r]{$\alpha=4.0$}}
\put(245,804){\makebox(0,0){$\triangle$}}
\put(251,804){\makebox(0,0){$\triangle$}}
\put(257,804){\makebox(0,0){$\triangle$}}
\put(306,804){\makebox(0,0){$\triangle$}}
\put(366,804){\makebox(0,0){$\triangle$}}
\put(487,804){\makebox(0,0){$\triangle$}}
\put(608,804){\makebox(0,0){$\triangle$}}
\put(729,803){\makebox(0,0){$\triangle$}}
\put(851,800){\makebox(0,0){$\triangle$}}
\put(972,794){\makebox(0,0){$\triangle$}}
\put(1093,784){\makebox(0,0){$\triangle$}}
\put(1214,768){\makebox(0,0){$\triangle$}}
\put(1335,744){\makebox(0,0){$\triangle$}}
\put(1359,739){\makebox(0,0){$\triangle$}}
\put(1371,737){\makebox(0,0){$\triangle$}}
\put(1383,733){\makebox(0,0){$\triangle$}}
\put(1395,730){\makebox(0,0){$\triangle$}}
\put(1408,727){\makebox(0,0){$\triangle$}}
\put(1420,723){\makebox(0,0){$\triangle$}}
\put(1432,719){\makebox(0,0){$\triangle$}}
\put(1456,711){\makebox(0,0){$\triangle$}}
\put(1389,171){\makebox(0,0){$\triangle$}}
\end{picture}

\caption{Dynamical fermion mass as a function of the temperature $T$ 
for $N=0$ at $p=0.1\Lambda$}
\end{center}
\end{figure}
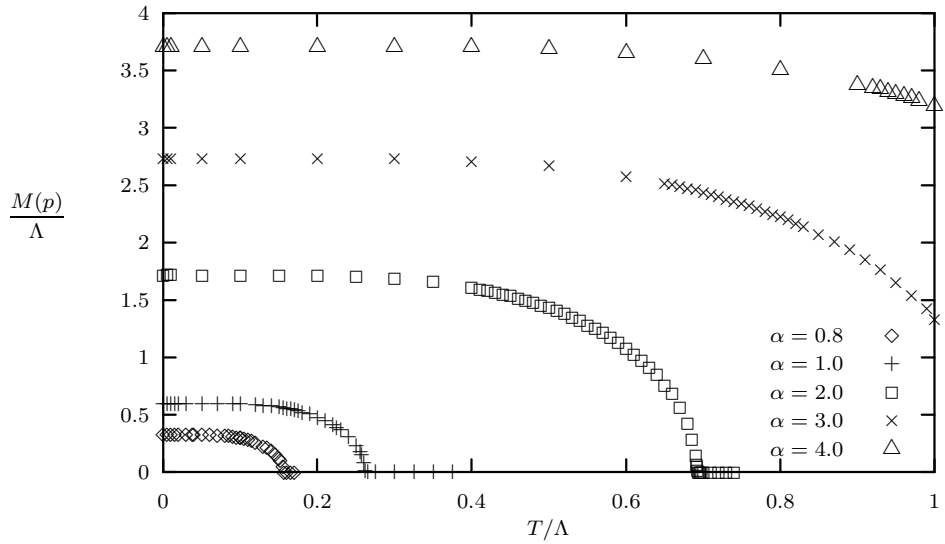

As is clearly seen in Fig. 3, the chiral phase transition is
of the second order since the fermion mass is generated at a critical
value of the coupling constant $\alpha$ without any discontinuity.
By picking up the critical value of $\alpha$ where the mass generation
occurs one can obtain the critical parameters $\alpha_c$ and $T_c$
respectively. If $T_c$ is plotted versus $\alpha_c$, we obtain the
phase diagram. We shall come back to this subject later in Section 6.
It should be noted here that the value of the critical coupling constant
$\alpha_c$ for $T=0$ is found to be $0.44477\pm 0.00053$ (See Appendix A).
This value differs from the
well-known value $\pi/3$. The reason for this is simple, i. e., 
we applied the instantaneous exchange approximation in our approach
which is not the good approximation in the low temperature region.
Thus we have to be careful in applying our method and consider
that our formulation is suitable rather in the high temperature region.

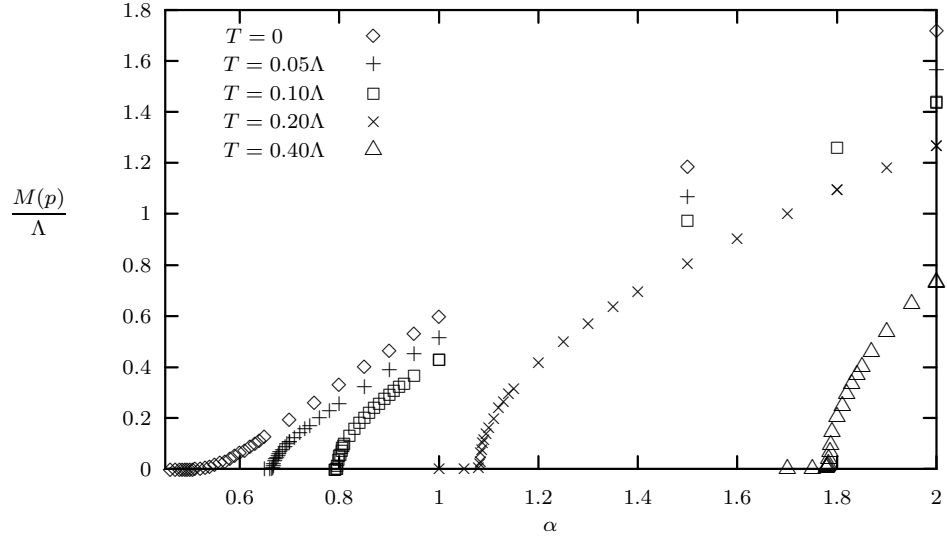
\begin{figure}
\begin{center}
\setlength{\unitlength}{0.240900pt}
\begin{picture}(1500,900)(0,0)
\footnotesize
\thicklines \path(245,135)(265,135)
\thicklines \path(1456,135)(1436,135)
\put(223,135){\makebox(0,0)[r]{0}}
\thicklines \path(245,215)(265,215)
\thicklines \path(1456,215)(1436,215)
\put(223,215){\makebox(0,0)[r]{0.2}}
\thicklines \path(245,295)(265,295)
\thicklines \path(1456,295)(1436,295)
\put(223,295){\makebox(0,0)[r]{0.4}}
\thicklines \path(245,375)(265,375)
\thicklines \path(1456,375)(1436,375)
\put(223,375){\makebox(0,0)[r]{0.6}}
\thicklines \path(245,455)(265,455)
\thicklines \path(1456,455)(1436,455)
\put(223,455){\makebox(0,0)[r]{0.8}}
\thicklines \path(245,536)(265,536)
\thicklines \path(1456,536)(1436,536)
\put(223,536){\makebox(0,0)[r]{1}}
\thicklines \path(245,616)(265,616)
\thicklines \path(1456,616)(1436,616)
\put(223,616){\makebox(0,0)[r]{1.2}}
\thicklines \path(245,696)(265,696)
\thicklines \path(1456,696)(1436,696)
\put(223,696){\makebox(0,0)[r]{1.4}}
\thicklines \path(245,776)(265,776)
\thicklines \path(1456,776)(1436,776)
\put(223,776){\makebox(0,0)[r]{1.6}}
\thicklines \path(245,856)(265,856)
\thicklines \path(1456,856)(1436,856)
\put(223,856){\makebox(0,0)[r]{1.8}}
\thicklines \path(362,135)(362,155)
\thicklines \path(362,856)(362,836)
\put(362,90){\makebox(0,0){0.6}}
\thicklines \path(518,135)(518,155)
\thicklines \path(518,856)(518,836)
\put(518,90){\makebox(0,0){0.8}}
\thicklines \path(675,135)(675,155)
\thicklines \path(675,856)(675,836)
\put(675,90){\makebox(0,0){1}}
\thicklines \path(831,135)(831,155)
\thicklines \path(831,856)(831,836)
\put(831,90){\makebox(0,0){1.2}}
\thicklines \path(987,135)(987,155)
\thicklines \path(987,856)(987,836)
\put(987,90){\makebox(0,0){1.4}}
\thicklines \path(1143,135)(1143,155)
\thicklines \path(1143,856)(1143,836)
\put(1143,90){\makebox(0,0){1.6}}
\thicklines \path(1300,135)(1300,155)
\thicklines \path(1300,856)(1300,836)
\put(1300,90){\makebox(0,0){1.8}}
\thicklines \path(1456,135)(1456,155)
\thicklines \path(1456,856)(1456,836)
\put(1456,90){\makebox(0,0){2}}
\thicklines \path(245,135)(1456,135)(1456,856)(245,856)(245,135)
\put(0,539){\makebox(0,0)[l]
{\shortstack{$\displaystyle\frac{M(p)}{\Lambda}$}}}
\put(850,45){\makebox(0,0){$\alpha$}}
\put(496,816){\makebox(0,0)[r]{$T=0$\hspace{4.6ex}}}
\put(253,135){\raisebox{-1.2pt}{\makebox(0,0){$\Diamond$}}}
\put(261,135){\raisebox{-1.2pt}{\makebox(0,0){$\Diamond$}}}
\put(268,135){\raisebox{-1.2pt}{\makebox(0,0){$\Diamond$}}}
\put(272,135){\raisebox{-1.2pt}{\makebox(0,0){$\Diamond$}}}
\put(276,135){\raisebox{-1.2pt}{\makebox(0,0){$\Diamond$}}}
\put(280,135){\raisebox{-1.2pt}{\makebox(0,0){$\Diamond$}}}
\put(284,135){\raisebox{-1.2pt}{\makebox(0,0){$\Diamond$}}}
\put(288,135){\raisebox{-1.2pt}{\makebox(0,0){$\Diamond$}}}
\put(292,136){\raisebox{-1.2pt}{\makebox(0,0){$\Diamond$}}}
\put(300,136){\raisebox{-1.2pt}{\makebox(0,0){$\Diamond$}}}
\put(308,138){\raisebox{-1.2pt}{\makebox(0,0){$\Diamond$}}}
\put(315,139){\raisebox{-1.2pt}{\makebox(0,0){$\Diamond$}}}
\put(323,142){\raisebox{-1.2pt}{\makebox(0,0){$\Diamond$}}}
\put(331,145){\raisebox{-1.2pt}{\makebox(0,0){$\Diamond$}}}
\put(339,148){\raisebox{-1.2pt}{\makebox(0,0){$\Diamond$}}}
\put(347,152){\raisebox{-1.2pt}{\makebox(0,0){$\Diamond$}}}
\put(354,157){\raisebox{-1.2pt}{\makebox(0,0){$\Diamond$}}}
\put(362,162){\raisebox{-1.2pt}{\makebox(0,0){$\Diamond$}}}
\put(370,165){\raisebox{-1.2pt}{\makebox(0,0){$\Diamond$}}}
\put(378,171){\raisebox{-1.2pt}{\makebox(0,0){$\Diamond$}}}
\put(386,176){\raisebox{-1.2pt}{\makebox(0,0){$\Diamond$}}}
\put(393,181){\raisebox{-1.2pt}{\makebox(0,0){$\Diamond$}}}
\put(401,186){\raisebox{-1.2pt}{\makebox(0,0){$\Diamond$}}}
\put(440,213){\raisebox{-1.2pt}{\makebox(0,0){$\Diamond$}}}
\put(479,240){\raisebox{-1.2pt}{\makebox(0,0){$\Diamond$}}}
\put(518,268){\raisebox{-1.2pt}{\makebox(0,0){$\Diamond$}}}
\put(558,296){\raisebox{-1.2pt}{\makebox(0,0){$\Diamond$}}}
\put(597,322){\raisebox{-1.2pt}{\makebox(0,0){$\Diamond$}}}
\put(636,349){\raisebox{-1.2pt}{\makebox(0,0){$\Diamond$}}}
\put(675,375){\raisebox{-1.2pt}{\makebox(0,0){$\Diamond$}}}
\put(1065,611){\raisebox{-1.2pt}{\makebox(0,0){$\Diamond$}}}
\put(1456,825){\raisebox{-1.2pt}{\makebox(0,0){$\Diamond$}}}
\put(572,816){\raisebox{-1.2pt}{\makebox(0,0){$\Diamond$}}}
\put(496,771){\makebox(0,0)[r]{$T=0.05\Lambda$}}
\put(401,135){\makebox(0,0){$+$}}
\put(409,135){\makebox(0,0){$+$}}
\put(412,135){\makebox(0,0){$+$}}
\put(414,137){\makebox(0,0){$+$}}
\put(415,139){\makebox(0,0){$+$}}
\put(415,142){\makebox(0,0){$+$}}
\put(417,148){\makebox(0,0){$+$}}
\put(418,150){\makebox(0,0){$+$}}
\put(420,153){\makebox(0,0){$+$}}
\put(422,156){\makebox(0,0){$+$}}
\put(423,158){\makebox(0,0){$+$}}
\put(425,161){\makebox(0,0){$+$}}
\put(429,165){\makebox(0,0){$+$}}
\put(433,170){\makebox(0,0){$+$}}
\put(436,174){\makebox(0,0){$+$}}
\put(440,178){\makebox(0,0){$+$}}
\put(440,178){\makebox(0,0){$+$}}
\put(448,184){\makebox(0,0){$+$}}
\put(456,191){\makebox(0,0){$+$}}
\put(464,198){\makebox(0,0){$+$}}
\put(472,203){\makebox(0,0){$+$}}
\put(487,215){\makebox(0,0){$+$}}
\put(503,226){\makebox(0,0){$+$}}
\put(518,237){\makebox(0,0){$+$}}
\put(558,264){\makebox(0,0){$+$}}
\put(597,291){\makebox(0,0){$+$}}
\put(636,315){\makebox(0,0){$+$}}
\put(675,340){\makebox(0,0){$+$}}
\put(1065,563){\makebox(0,0){$+$}}
\put(1456,762){\makebox(0,0){$+$}}
\put(572,771){\makebox(0,0){$+$}}
\put(496,726){\makebox(0,0)[r]{$T=0.10\Lambda$}}
\put(511,135){\raisebox{-1.2pt}{\makebox(0,0){$\Box$}}}
\put(512,135){\raisebox{-1.2pt}{\makebox(0,0){$\Box$}}}
\put(514,136){\raisebox{-1.2pt}{\makebox(0,0){$\Box$}}}
\put(515,136){\raisebox{-1.2pt}{\makebox(0,0){$\Box$}}}
\put(515,141){\raisebox{-1.2pt}{\makebox(0,0){$\Box$}}}
\put(517,151){\raisebox{-1.2pt}{\makebox(0,0){$\Box$}}}
\put(518,157){\raisebox{-1.2pt}{\makebox(0,0){$\Box$}}}
\put(520,159){\raisebox{-1.2pt}{\makebox(0,0){$\Box$}}}
\put(522,165){\raisebox{-1.2pt}{\makebox(0,0){$\Box$}}}
\put(523,169){\raisebox{-1.2pt}{\makebox(0,0){$\Box$}}}
\put(525,173){\raisebox{-1.2pt}{\makebox(0,0){$\Box$}}}
\put(526,176){\raisebox{-1.2pt}{\makebox(0,0){$\Box$}}}
\put(534,188){\raisebox{-1.2pt}{\makebox(0,0){$\Box$}}}
\put(542,199){\raisebox{-1.2pt}{\makebox(0,0){$\Box$}}}
\put(550,208){\raisebox{-1.2pt}{\makebox(0,0){$\Box$}}}
\put(558,217){\raisebox{-1.2pt}{\makebox(0,0){$\Box$}}}
\put(565,225){\raisebox{-1.2pt}{\makebox(0,0){$\Box$}}}
\put(573,232){\raisebox{-1.2pt}{\makebox(0,0){$\Box$}}}
\put(581,239){\raisebox{-1.2pt}{\makebox(0,0){$\Box$}}}
\put(589,246){\raisebox{-1.2pt}{\makebox(0,0){$\Box$}}}
\put(597,252){\raisebox{-1.2pt}{\makebox(0,0){$\Box$}}}
\put(604,259){\raisebox{-1.2pt}{\makebox(0,0){$\Box$}}}
\put(612,265){\raisebox{-1.2pt}{\makebox(0,0){$\Box$}}}
\put(620,270){\raisebox{-1.2pt}{\makebox(0,0){$\Box$}}}
\put(636,282){\raisebox{-1.2pt}{\makebox(0,0){$\Box$}}}
\put(675,308){\raisebox{-1.2pt}{\makebox(0,0){$\Box$}}}
\put(675,308){\raisebox{-1.2pt}{\makebox(0,0){$\Box$}}}
\put(1065,526){\raisebox{-1.2pt}{\makebox(0,0){$\Box$}}}
\put(1300,640){\raisebox{-1.2pt}{\makebox(0,0){$\Box$}}}
\put(1456,712){\raisebox{-1.2pt}{\makebox(0,0){$\Box$}}}
\put(1456,713){\raisebox{-1.2pt}{\makebox(0,0){$\Box$}}}
\put(572,726){\raisebox{-1.2pt}{\makebox(0,0){$\Box$}}}
\put(496,681){\makebox(0,0)[r]{$T=0.20\Lambda$}}
\put(675,135){\makebox(0,0){$\times$}}
\put(714,135){\makebox(0,0){$\times$}}
\put(737,137){\makebox(0,0){$\times$}}
\put(739,142){\makebox(0,0){$\times$}}
\put(740,146){\makebox(0,0){$\times$}}
\put(740,160){\makebox(0,0){$\times$}}
\put(742,165){\makebox(0,0){$\times$}}
\put(743,175){\makebox(0,0){$\times$}}
\put(745,180){\makebox(0,0){$\times$}}
\put(749,190){\makebox(0,0){$\times$}}
\put(753,199){\makebox(0,0){$\times$}}
\put(761,214){\makebox(0,0){$\times$}}
\put(768,230){\makebox(0,0){$\times$}}
\put(776,241){\makebox(0,0){$\times$}}
\put(784,252){\makebox(0,0){$\times$}}
\put(792,260){\makebox(0,0){$\times$}}
\put(831,302){\makebox(0,0){$\times$}}
\put(870,334){\makebox(0,0){$\times$}}
\put(909,363){\makebox(0,0){$\times$}}
\put(948,389){\makebox(0,0){$\times$}}
\put(987,413){\makebox(0,0){$\times$}}
\put(1065,457){\makebox(0,0){$\times$}}
\put(1143,497){\makebox(0,0){$\times$}}
\put(1222,536){\makebox(0,0){$\times$}}
\put(1300,573){\makebox(0,0){$\times$}}
\put(1300,573){\makebox(0,0){$\times$}}
\put(1378,608){\makebox(0,0){$\times$}}
\put(1456,642){\makebox(0,0){$\times$}}
\put(1456,643){\makebox(0,0){$\times$}}
\put(572,681){\makebox(0,0){$\times$}}
\put(496,636){\makebox(0,0)[r]{$T=0.40\Lambda$}}
\put(1222,135){\makebox(0,0){$\triangle$}}
\put(1261,135){\makebox(0,0){$\triangle$}}
\put(1283,137){\makebox(0,0){$\triangle$}}
\put(1284,138){\makebox(0,0){$\triangle$}}
\put(1285,140){\makebox(0,0){$\triangle$}}
\put(1286,142){\makebox(0,0){$\triangle$}}
\put(1287,151){\makebox(0,0){$\triangle$}}
\put(1289,162){\makebox(0,0){$\triangle$}}
\put(1290,172){\makebox(0,0){$\triangle$}}
\put(1292,193){\makebox(0,0){$\triangle$}}
\put(1300,217){\makebox(0,0){$\triangle$}}
\put(1308,234){\makebox(0,0){$\triangle$}}
\put(1315,253){\makebox(0,0){$\triangle$}}
\put(1323,268){\makebox(0,0){$\triangle$}}
\put(1331,282){\makebox(0,0){$\triangle$}}
\put(1339,296){\makebox(0,0){$\triangle$}}
\put(1354,318){\makebox(0,0){$\triangle$}}
\put(1378,350){\makebox(0,0){$\triangle$}}
\put(1417,394){\makebox(0,0){$\triangle$}}
\put(1456,427){\makebox(0,0){$\triangle$}}
\put(1456,428){\makebox(0,0){$\triangle$}}
\put(572,636){\makebox(0,0){$\triangle$}}
\end{picture}

\caption{Dynamical fermion mass as a function of the coupling constant 
$\alpha$ for $N=1$ at $p=0.1\Lambda$}
\end{center}
\end{figure}
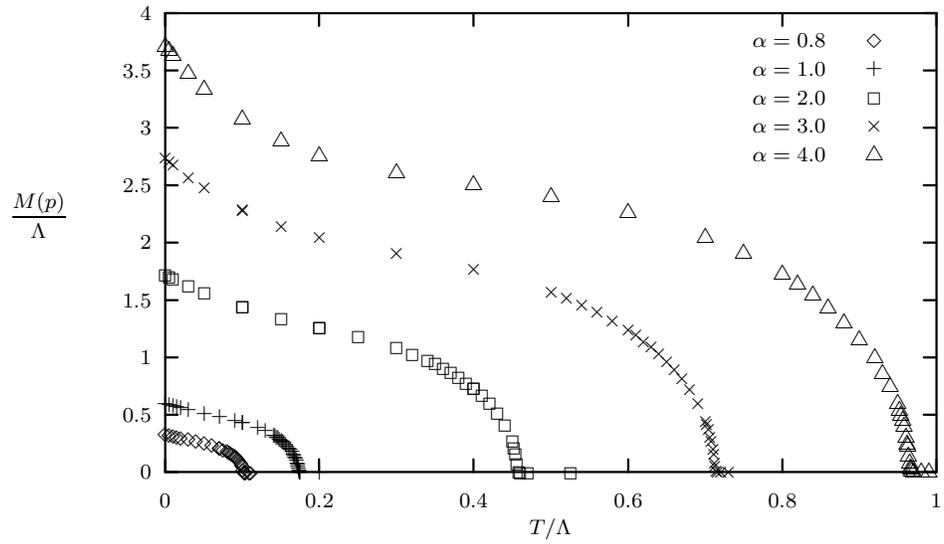
\begin{figure}
\begin{center}
\setlength{\unitlength}{0.240900pt}
\begin{picture}(1500,900)(0,0)
\footnotesize
\thicklines \path(245,135)(265,135)
\thicklines \path(1456,135)(1436,135)
\put(223,135){\makebox(0,0)[r]{0}}
\thicklines \path(245,225)(265,225)
\thicklines \path(1456,225)(1436,225)
\put(223,225){\makebox(0,0)[r]{0.5}}
\thicklines \path(245,315)(265,315)
\thicklines \path(1456,315)(1436,315)
\put(223,315){\makebox(0,0)[r]{1}}
\thicklines \path(245,405)(265,405)
\thicklines \path(1456,405)(1436,405)
\put(223,405){\makebox(0,0)[r]{1.5}}
\thicklines \path(245,496)(265,496)
\thicklines \path(1456,496)(1436,496)
\put(223,496){\makebox(0,0)[r]{2}}
\thicklines \path(245,586)(265,586)
\thicklines \path(1456,586)(1436,586)
\put(223,586){\makebox(0,0)[r]{2.5}}
\thicklines \path(245,676)(265,676)
\thicklines \path(1456,676)(1436,676)
\put(223,676){\makebox(0,0)[r]{3}}
\thicklines \path(245,766)(265,766)
\thicklines \path(1456,766)(1436,766)
\put(223,766){\makebox(0,0)[r]{3.5}}
\thicklines \path(245,856)(265,856)
\thicklines \path(1456,856)(1436,856)
\put(223,856){\makebox(0,0)[r]{4}}
\thicklines \path(245,135)(245,155)
\thicklines \path(245,856)(245,836)
\put(245,90){\makebox(0,0){0}}
\thicklines \path(487,135)(487,155)
\thicklines \path(487,856)(487,836)
\put(487,90){\makebox(0,0){0.2}}
\thicklines \path(729,135)(729,155)
\thicklines \path(729,856)(729,836)
\put(729,90){\makebox(0,0){0.4}}
\thicklines \path(972,135)(972,155)
\thicklines \path(972,856)(972,836)
\put(972,90){\makebox(0,0){0.6}}
\thicklines \path(1214,135)(1214,155)
\thicklines \path(1214,856)(1214,836)
\put(1214,90){\makebox(0,0){0.8}}
\thicklines \path(1456,135)(1456,155)
\thicklines \path(1456,856)(1456,836)
\put(1456,90){\makebox(0,0){1}}
\thicklines \path(245,135)(1456,135)(1456,856)(245,856)(245,135)
\put(0,539){\makebox(0,0)[l]
{\shortstack{$\displaystyle\frac{M(p)}{\Lambda}$}}}
\put(850,45){\makebox(0,0){$T/\Lambda$}}
\put(1282,814){\makebox(0,0)[r]{$\alpha=0.8$}}
\put(245,195){\raisebox{-1.2pt}{\makebox(0,0){$\Diamond$}}}
\put(251,194){\raisebox{-1.2pt}{\makebox(0,0){$\Diamond$}}}
\put(257,192){\raisebox{-1.2pt}{\makebox(0,0){$\Diamond$}}}
\put(263,191){\raisebox{-1.2pt}{\makebox(0,0){$\Diamond$}}}
\put(269,189){\raisebox{-1.2pt}{\makebox(0,0){$\Diamond$}}}
\put(281,187){\raisebox{-1.2pt}{\makebox(0,0){$\Diamond$}}}
\put(293,184){\raisebox{-1.2pt}{\makebox(0,0){$\Diamond$}}}
\put(306,181){\raisebox{-1.2pt}{\makebox(0,0){$\Diamond$}}}
\put(318,178){\raisebox{-1.2pt}{\makebox(0,0){$\Diamond$}}}
\put(330,174){\raisebox{-1.2pt}{\makebox(0,0){$\Diamond$}}}
\put(330,174){\raisebox{-1.2pt}{\makebox(0,0){$\Diamond$}}}
\put(336,171){\raisebox{-1.2pt}{\makebox(0,0){$\Diamond$}}}
\put(342,169){\raisebox{-1.2pt}{\makebox(0,0){$\Diamond$}}}
\put(344,167){\raisebox{-1.2pt}{\makebox(0,0){$\Diamond$}}}
\put(347,166){\raisebox{-1.2pt}{\makebox(0,0){$\Diamond$}}}
\put(349,164){\raisebox{-1.2pt}{\makebox(0,0){$\Diamond$}}}
\put(352,162){\raisebox{-1.2pt}{\makebox(0,0){$\Diamond$}}}
\put(354,161){\raisebox{-1.2pt}{\makebox(0,0){$\Diamond$}}}
\put(356,159){\raisebox{-1.2pt}{\makebox(0,0){$\Diamond$}}}
\put(359,156){\raisebox{-1.2pt}{\makebox(0,0){$\Diamond$}}}
\put(361,153){\raisebox{-1.2pt}{\makebox(0,0){$\Diamond$}}}
\put(364,150){\raisebox{-1.2pt}{\makebox(0,0){$\Diamond$}}}
\put(365,147){\raisebox{-1.2pt}{\makebox(0,0){$\Diamond$}}}
\put(366,145){\raisebox{-1.2pt}{\makebox(0,0){$\Diamond$}}}
\put(367,139){\raisebox{-1.2pt}{\makebox(0,0){$\Diamond$}}}
\put(369,136){\raisebox{-1.2pt}{\makebox(0,0){$\Diamond$}}}
\put(370,135){\raisebox{-1.2pt}{\makebox(0,0){$\Diamond$}}}
\put(371,135){\raisebox{-1.2pt}{\makebox(0,0){$\Diamond$}}}
\put(377,135){\raisebox{-1.2pt}{\makebox(0,0){$\Diamond$}}}
\put(379,135){\raisebox{-1.2pt}{\makebox(0,0){$\Diamond$}}}
\put(1358,814){\raisebox{-1.2pt}{\makebox(0,0){$\Diamond$}}}
\put(1282,769){\makebox(0,0)[r]{$\alpha=1.0$}}
\put(245,243){\makebox(0,0){$+$}}
\put(251,241){\makebox(0,0){$+$}}
\put(257,240){\makebox(0,0){$+$}}
\put(263,238){\makebox(0,0){$+$}}
\put(269,236){\makebox(0,0){$+$}}
\put(281,233){\makebox(0,0){$+$}}
\put(306,227){\makebox(0,0){$+$}}
\put(330,222){\makebox(0,0){$+$}}
\put(354,216){\makebox(0,0){$+$}}
\put(366,213){\makebox(0,0){$+$}}
\put(366,213){\makebox(0,0){$+$}}
\put(390,205){\makebox(0,0){$+$}}
\put(402,200){\makebox(0,0){$+$}}
\put(415,193){\makebox(0,0){$+$}}
\put(417,192){\makebox(0,0){$+$}}
\put(419,190){\makebox(0,0){$+$}}
\put(422,189){\makebox(0,0){$+$}}
\put(424,187){\makebox(0,0){$+$}}
\put(427,185){\makebox(0,0){$+$}}
\put(429,183){\makebox(0,0){$+$}}
\put(431,181){\makebox(0,0){$+$}}
\put(434,178){\makebox(0,0){$+$}}
\put(436,176){\makebox(0,0){$+$}}
\put(439,173){\makebox(0,0){$+$}}
\put(441,170){\makebox(0,0){$+$}}
\put(444,167){\makebox(0,0){$+$}}
\put(446,163){\makebox(0,0){$+$}}
\put(447,161){\makebox(0,0){$+$}}
\put(448,159){\makebox(0,0){$+$}}
\put(450,156){\makebox(0,0){$+$}}
\put(451,153){\makebox(0,0){$+$}}
\put(452,149){\makebox(0,0){$+$}}
\put(453,147){\makebox(0,0){$+$}}
\put(453,144){\makebox(0,0){$+$}}
\put(454,141){\makebox(0,0){$+$}}
\put(455,138){\makebox(0,0){$+$}}
\put(455,136){\makebox(0,0){$+$}}
\put(456,135){\makebox(0,0){$+$}}
\put(456,135){\makebox(0,0){$+$}}
\put(487,135){\makebox(0,0){$+$}}
\put(1358,769){\makebox(0,0){$+$}}
\put(1282,724){\makebox(0,0)[r]{$\alpha=2.0$}}
\put(245,445){\raisebox{-1.2pt}{\makebox(0,0){$\Box$}}}
\put(251,443){\raisebox{-1.2pt}{\makebox(0,0){$\Box$}}}
\put(257,440){\raisebox{-1.2pt}{\makebox(0,0){$\Box$}}}
\put(281,428){\raisebox{-1.2pt}{\makebox(0,0){$\Box$}}}
\put(306,417){\raisebox{-1.2pt}{\makebox(0,0){$\Box$}}}
\put(366,395){\raisebox{-1.2pt}{\makebox(0,0){$\Box$}}}
\put(366,395){\raisebox{-1.2pt}{\makebox(0,0){$\Box$}}}
\put(427,377){\raisebox{-1.2pt}{\makebox(0,0){$\Box$}}}
\put(487,363){\raisebox{-1.2pt}{\makebox(0,0){$\Box$}}}
\put(487,363){\raisebox{-1.2pt}{\makebox(0,0){$\Box$}}}
\put(548,348){\raisebox{-1.2pt}{\makebox(0,0){$\Box$}}}
\put(608,331){\raisebox{-1.2pt}{\makebox(0,0){$\Box$}}}
\put(633,321){\raisebox{-1.2pt}{\makebox(0,0){$\Box$}}}
\put(657,311){\raisebox{-1.2pt}{\makebox(0,0){$\Box$}}}
\put(669,306){\raisebox{-1.2pt}{\makebox(0,0){$\Box$}}}
\put(681,299){\raisebox{-1.2pt}{\makebox(0,0){$\Box$}}}
\put(693,292){\raisebox{-1.2pt}{\makebox(0,0){$\Box$}}}
\put(705,285){\raisebox{-1.2pt}{\makebox(0,0){$\Box$}}}
\put(717,276){\raisebox{-1.2pt}{\makebox(0,0){$\Box$}}}
\put(729,267){\raisebox{-1.2pt}{\makebox(0,0){$\Box$}}}
\put(729,267){\raisebox{-1.2pt}{\makebox(0,0){$\Box$}}}
\put(742,256){\raisebox{-1.2pt}{\makebox(0,0){$\Box$}}}
\put(754,244){\raisebox{-1.2pt}{\makebox(0,0){$\Box$}}}
\put(766,229){\raisebox{-1.2pt}{\makebox(0,0){$\Box$}}}
\put(778,209){\raisebox{-1.2pt}{\makebox(0,0){$\Box$}}}
\put(790,184){\raisebox{-1.2pt}{\makebox(0,0){$\Box$}}}
\put(792,174){\raisebox{-1.2pt}{\makebox(0,0){$\Box$}}}
\put(795,164){\raisebox{-1.2pt}{\makebox(0,0){$\Box$}}}
\put(797,149){\raisebox{-1.2pt}{\makebox(0,0){$\Box$}}}
\put(800,136){\raisebox{-1.2pt}{\makebox(0,0){$\Box$}}}
\put(802,135){\raisebox{-1.2pt}{\makebox(0,0){$\Box$}}}
\put(814,135){\raisebox{-1.2pt}{\makebox(0,0){$\Box$}}}
\put(881,135){\raisebox{-1.2pt}{\makebox(0,0){$\Box$}}}
\put(1358,724){\raisebox{-1.2pt}{\makebox(0,0){$\Box$}}}
\put(1282,679){\makebox(0,0)[r]{$\alpha=3.0$}}
\put(245,628){\makebox(0,0){$\times$}}
\put(251,622){\makebox(0,0){$\times$}}
\put(257,618){\makebox(0,0){$\times$}}
\put(281,598){\makebox(0,0){$\times$}}
\put(306,581){\makebox(0,0){$\times$}}
\put(366,546){\makebox(0,0){$\times$}}
\put(366,547){\makebox(0,0){$\times$}}
\put(427,521){\makebox(0,0){$\times$}}
\put(487,503){\makebox(0,0){$\times$}}
\put(608,478){\makebox(0,0){$\times$}}
\put(729,453){\makebox(0,0){$\times$}}
\put(851,418){\makebox(0,0){$\times$}}
\put(875,408){\makebox(0,0){$\times$}}
\put(899,398){\makebox(0,0){$\times$}}
\put(923,386){\makebox(0,0){$\times$}}
\put(947,373){\makebox(0,0){$\times$}}
\put(972,359){\makebox(0,0){$\times$}}
\put(984,350){\makebox(0,0){$\times$}}
\put(996,340){\makebox(0,0){$\times$}}
\put(1008,331){\makebox(0,0){$\times$}}
\put(1020,321){\makebox(0,0){$\times$}}
\put(1032,308){\makebox(0,0){$\times$}}
\put(1044,296){\makebox(0,0){$\times$}}
\put(1056,282){\makebox(0,0){$\times$}}
\put(1068,264){\makebox(0,0){$\times$}}
\put(1081,243){\makebox(0,0){$\times$}}
\put(1093,214){\makebox(0,0){$\times$}}
\put(1095,210){\makebox(0,0){$\times$}}
\put(1098,202){\makebox(0,0){$\times$}}
\put(1100,189){\makebox(0,0){$\times$}}
\put(1102,183){\makebox(0,0){$\times$}}
\put(1105,171){\makebox(0,0){$\times$}}
\put(1106,163){\makebox(0,0){$\times$}}
\put(1107,149){\makebox(0,0){$\times$}}
\put(1108,144){\makebox(0,0){$\times$}}
\put(1110,137){\makebox(0,0){$\times$}}
\put(1112,135){\makebox(0,0){$\times$}}
\put(1117,135){\makebox(0,0){$\times$}}
\put(1129,135){\makebox(0,0){$\times$}}
\put(1358,679){\makebox(0,0){$\times$}}
\put(1282,634){\makebox(0,0)[r]{$\alpha=4.0$}}
\put(245,804){\makebox(0,0){$\triangle$}}
\put(251,797){\makebox(0,0){$\triangle$}}
\put(257,789){\makebox(0,0){$\triangle$}}
\put(281,762){\makebox(0,0){$\triangle$}}
\put(306,737){\makebox(0,0){$\triangle$}}
\put(366,689){\makebox(0,0){$\triangle$}}
\put(427,655){\makebox(0,0){$\triangle$}}
\put(487,632){\makebox(0,0){$\triangle$}}
\put(608,605){\makebox(0,0){$\triangle$}}
\put(729,587){\makebox(0,0){$\triangle$}}
\put(851,568){\makebox(0,0){$\triangle$}}
\put(972,542){\makebox(0,0){$\triangle$}}
\put(1093,504){\makebox(0,0){$\triangle$}}
\put(1153,478){\makebox(0,0){$\triangle$}}
\put(1214,446){\makebox(0,0){$\triangle$}}
\put(1238,430){\makebox(0,0){$\triangle$}}
\put(1262,413){\makebox(0,0){$\triangle$}}
\put(1286,393){\makebox(0,0){$\triangle$}}
\put(1311,369){\makebox(0,0){$\triangle$}}
\put(1335,343){\makebox(0,0){$\triangle$}}
\put(1359,314){\makebox(0,0){$\triangle$}}
\put(1371,289){\makebox(0,0){$\triangle$}}
\put(1383,269){\makebox(0,0){$\triangle$}}
\put(1395,242){\makebox(0,0){$\triangle$}}
\put(1398,231){\makebox(0,0){$\triangle$}}
\put(1400,224){\makebox(0,0){$\triangle$}}
\put(1403,216){\makebox(0,0){$\triangle$}}
\put(1405,206){\makebox(0,0){$\triangle$}}
\put(1408,190){\makebox(0,0){$\triangle$}}
\put(1410,179){\makebox(0,0){$\triangle$}}
\put(1411,175){\makebox(0,0){$\triangle$}}
\put(1412,160){\makebox(0,0){$\triangle$}}
\put(1414,149){\makebox(0,0){$\triangle$}}
\put(1415,141){\makebox(0,0){$\triangle$}}
\put(1416,140){\makebox(0,0){$\triangle$}}
\put(1417,136){\makebox(0,0){$\triangle$}}
\put(1420,135){\makebox(0,0){$\triangle$}}
\put(1432,135){\makebox(0,0){$\triangle$}}
\put(1444,135){\makebox(0,0){$\triangle$}}
\put(1358,634){\makebox(0,0){$\triangle$}}
\end{picture}

\caption{Dynamical fermion mass as a function of the temperature $T$ 
for $N=1$ at $p=0.1\Lambda$}
\end{center}
\end{figure}

We can also observe the behavior of the mass function $M(p)$ at
some fixed value of $p$ as a function of temperature $T$. The $T$
dependence of $M(p)$ with $p=0.1\Lambda$ is given in Fig. 4 for
fixed $\alpha$.
Here again we clearly observe the second order chiral phase transition
in Fig. 4.

If the thermal photon mass is included ($N=1$), the essential feature
of the behavior of the mass function is more or less the same as in
the case without the thermal photon mass. We present the $\alpha$
and $T$ dependence of the mass function $M(p)$ are presented in
Figs. 5 and 6.

We recognize that the behavior of the mass function in the low
temperature region as seen in Figs. 4 and 6 is significantly different
from the one in the case without the thermal photon mass. The similar 
behavior is observed in the case of $N=3$ although the result is not shown 
in figures.

\section{Critical curve and critical exponents}

By observing the figures obtained in the last section we can directly
draw the critical curve on the $T-\alpha$ plane. In fact, for the case
with $N=0$ we pick out the values of $\alpha$ where the mass function
vanishes and plot those values as a function of temperature. The
resulting curve is the critical curve for the case with $N=0$.
In order to perform this procedure in a more systematic way and
to obtain critical exponents simultaneously we apply the following
method.

We fit the function $M(p)$ with $n$ data points $M_i, \alpha_i$ and $T_i$
with $i=1,\cdots n$ near the critical point by assuming the functional form
of the function $M(p)$ such that
\begin{equation}
M(p)=e^{C_{T}}(\alpha-\alpha_{c})^{\nu} ,
\end{equation}
for fixed temperature $T$ and
\begin{equation}
M(p)=e^{C_{\alpha}}(T_{c}-T)^{\eta} ,
\end{equation}
for fixed coupling constant $\alpha$ respectively.
Here $C_{T}, C_{\alpha}, \alpha_{c}, T_{c}, \nu$ and $\eta$ are
adjustable parameters with $\alpha_{c}$ and $T_{c}$ corresponding to
the critical coupling constant and critical temperature respectively
and $\nu$ and $\eta$ designate the critical exponent.
To estimate the values of $C_{T}, C_{\alpha}, \alpha_{c},$ $T_{c},$ 
$\nu$ and $\eta$ we adopt the familiar least-squares fit. We choose
$n$ data points near the critical point and minimize the quantities
\begin{equation}
\sum^{n}_{i=1}[\ln M_i -(\nu\ln(T_{c}-T_i)+C_{T})]^2 ,
\end{equation}
with temperature $T$ fixed and
\begin{equation}
\sum^{n}_{i=1}[\ln M_i -(\eta\ln(\alpha_i-\alpha_{c})+C_{\alpha})]^2 ,
\end{equation}
with coupling constant $\alpha$ fixed.

\begin{figure}
\begin{center}
\setlength{\unitlength}{0.240900pt}
\begin{picture}(1500,900)(0,0)
\footnotesize
\thicklines \path(155,135)(175,135)
\thicklines \path(1456,135)(1436,135)
\put(133,135){\makebox(0,0)[r]{0}}
\thicklines \path(155,217)(175,217)
\thicklines \path(1456,217)(1436,217)
\put(133,217){\makebox(0,0)[r]{0.5}}
\thicklines \path(155,299)(175,299)
\thicklines \path(1456,299)(1436,299)
\put(133,299){\makebox(0,0)[r]{1}}
\thicklines \path(155,381)(175,381)
\thicklines \path(1456,381)(1436,381)
\put(133,381){\makebox(0,0)[r]{1.5}}
\thicklines \path(155,463)(175,463)
\thicklines \path(1456,463)(1436,463)
\put(133,463){\makebox(0,0)[r]{2}}
\thicklines \path(155,545)(175,545)
\thicklines \path(1456,545)(1436,545)
\put(133,545){\makebox(0,0)[r]{2.5}}
\thicklines \path(155,627)(175,627)
\thicklines \path(1456,627)(1436,627)
\put(133,627){\makebox(0,0)[r]{3}}
\thicklines \path(155,709)(175,709)
\thicklines \path(1456,709)(1436,709)
\put(133,709){\makebox(0,0)[r]{3.5}}
\thicklines \path(155,790)(175,790)
\thicklines \path(1456,790)(1436,790)
\put(133,790){\makebox(0,0)[r]{4}}
\thicklines \path(155,135)(155,155)
\thicklines \path(155,856)(155,836)
\put(155,90){\makebox(0,0){0}}
\thicklines \path(387,135)(387,155)
\thicklines \path(387,856)(387,836)
\put(387,90){\makebox(0,0){0.2}}
\thicklines \path(620,135)(620,155)
\thicklines \path(620,856)(620,836)
\put(620,90){\makebox(0,0){0.4}}
\thicklines \path(852,135)(852,155)
\thicklines \path(852,856)(852,836)
\put(852,90){\makebox(0,0){0.6}}
\thicklines \path(1084,135)(1084,155)
\thicklines \path(1084,856)(1084,836)
\put(1084,90){\makebox(0,0){0.8}}
\thicklines \path(1317,135)(1317,155)
\thicklines \path(1317,856)(1317,836)
\put(1317,90){\makebox(0,0){1}}
\thicklines \path(155,135)(1456,135)(1456,856)(155,856)(155,135)
\put(0,495){\makebox(0,0)[l]{\shortstack{$\alpha_c$}}}
\put(805,45){\makebox(0,0){$T_c/\Lambda$}}
\put(365,807){\makebox(0,0)[r]{$N=0$}}
\thinlines \path(387,807)(495,807)
\thinlines \path(155,209)(155,209)(167,221)(204,233)(213,236)(271,250)
(336,266)(387,279)(458,299)(620,348)(852,425)(958,463)(1084,508)
(1317,592)(1410,627)(1456,644)
\put(155,209){\raisebox{-1.2pt}{\makebox(0,0){$\Diamond$}}}
\put(167,221){\raisebox{-1.2pt}{\makebox(0,0){$\Diamond$}}}
\put(204,233){\raisebox{-1.2pt}{\makebox(0,0){$\Diamond$}}}
\put(213,236){\raisebox{-1.2pt}{\makebox(0,0){$\Diamond$}}}
\put(271,250){\raisebox{-1.2pt}{\makebox(0,0){$\Diamond$}}}
\put(336,266){\raisebox{-1.2pt}{\makebox(0,0){$\Diamond$}}}
\put(387,279){\raisebox{-1.2pt}{\makebox(0,0){$\Diamond$}}}
\put(458,299){\raisebox{-1.2pt}{\makebox(0,0){$\Diamond$}}}
\put(620,348){\raisebox{-1.2pt}{\makebox(0,0){$\Diamond$}}}
\put(852,425){\raisebox{-1.2pt}{\makebox(0,0){$\Diamond$}}}
\put(958,463){\raisebox{-1.2pt}{\makebox(0,0){$\Diamond$}}}
\put(1084,508){\raisebox{-1.2pt}{\makebox(0,0){$\Diamond$}}}
\put(1317,592){\raisebox{-1.2pt}{\makebox(0,0){$\Diamond$}}}
\put(1410,627){\raisebox{-1.2pt}{\makebox(0,0){$\Diamond$}}}
\put(441,807){\raisebox{-1.2pt}{\makebox(0,0){$\Diamond$}}}
\put(365,762){\makebox(0,0)[r]{$N=1$}}
\thinlines \path(387,762)(495,762)
\thinlines \path(155,209)(155,209)(167,224)(185,233)(213,244)(236,252)
(271,265)(272,266)(318,283)(356,299)(387,312)(503,367)(620,427)(685,463)
(736,489)(852,554)(982,627)(1084,683)(1200,748)(1276,790)(1317,812)
\put(155,209){\makebox(0,0){$+$}}
\put(167,224){\makebox(0,0){$+$}}
\put(185,233){\makebox(0,0){$+$}}
\put(213,244){\makebox(0,0){$+$}}
\put(236,252){\makebox(0,0){$+$}}
\put(271,265){\makebox(0,0){$+$}}
\put(272,266){\makebox(0,0){$+$}}
\put(318,283){\makebox(0,0){$+$}}
\put(356,299){\makebox(0,0){$+$}}
\put(387,312){\makebox(0,0){$+$}}
\put(503,367){\makebox(0,0){$+$}}
\put(620,427){\makebox(0,0){$+$}}
\put(685,463){\makebox(0,0){$+$}}
\put(736,489){\makebox(0,0){$+$}}
\put(852,554){\makebox(0,0){$+$}}
\put(982,627){\makebox(0,0){$+$}}
\put(1084,683){\makebox(0,0){$+$}}
\put(1200,748){\makebox(0,0){$+$}}
\put(1276,790){\makebox(0,0){$+$}}
\put(1317,812){\makebox(0,0){$+$}}
\put(441,762){\makebox(0,0){$+$}}
\put(365,717){\makebox(0,0)[r]{$N=3$}}
\thinlines \path(387,717)(495,717)
\thinlines \path(155,209)(155,209)(167,226)(178,233)(213,249)(251,266)
(271,274)(326,299)(387,327)(620,442)(658,463)(852,565)(968,627)
(1084,690)(1268,790)(1317,818)(1387,856)
\put(155,209){\raisebox{-1.2pt}{\makebox(0,0){$\Box$}}}
\put(167,226){\raisebox{-1.2pt}{\makebox(0,0){$\Box$}}}
\put(178,233){\raisebox{-1.2pt}{\makebox(0,0){$\Box$}}}
\put(213,249){\raisebox{-1.2pt}{\makebox(0,0){$\Box$}}}
\put(251,266){\raisebox{-1.2pt}{\makebox(0,0){$\Box$}}}
\put(271,274){\raisebox{-1.2pt}{\makebox(0,0){$\Box$}}}
\put(326,299){\raisebox{-1.2pt}{\makebox(0,0){$\Box$}}}
\put(387,327){\raisebox{-1.2pt}{\makebox(0,0){$\Box$}}}
\put(620,442){\raisebox{-1.2pt}{\makebox(0,0){$\Box$}}}
\put(658,463){\raisebox{-1.2pt}{\makebox(0,0){$\Box$}}}
\put(852,565){\raisebox{-1.2pt}{\makebox(0,0){$\Box$}}}
\put(968,627){\raisebox{-1.2pt}{\makebox(0,0){$\Box$}}}
\put(1084,690){\raisebox{-1.2pt}{\makebox(0,0){$\Box$}}}
\put(1268,790){\raisebox{-1.2pt}{\makebox(0,0){$\Box$}}}
\put(1317,818){\raisebox{-1.2pt}{\makebox(0,0){$\Box$}}}
\put(441,717){\raisebox{-1.2pt}{\makebox(0,0){$\Box$}}}
\put(365,672){\makebox(0,0)[r]{$N=\infty$}}
\thinlines \path(387,672)(495,672)
\thinlines \path(155,246)(155,246)(167,264)(204,282)(213,286)(271,308)
(336,332)(387,351)(458,381)(620,455)(852,571)(958,627)(1084,694)
(1317,820)(1380,856)
\put(155,246){\makebox(0,0){$\times$}}
\put(167,264){\makebox(0,0){$\times$}}
\put(204,282){\makebox(0,0){$\times$}}
\put(213,286){\makebox(0,0){$\times$}}
\put(271,308){\makebox(0,0){$\times$}}
\put(336,332){\makebox(0,0){$\times$}}
\put(387,351){\makebox(0,0){$\times$}}
\put(458,381){\makebox(0,0){$\times$}}
\put(620,455){\makebox(0,0){$\times$}}
\put(852,571){\makebox(0,0){$\times$}}
\put(958,627){\makebox(0,0){$\times$}}
\put(1084,694){\makebox(0,0){$\times$}}
\put(1317,820){\makebox(0,0){$\times$}}
\put(441,672){\makebox(0,0){$\times$}}
\end{picture}

\caption{Critical curves for $N=0, N=1, N=3$ and $N=\infty$
at $p=0.1\Lambda$.}
\end{center}
\end{figure}
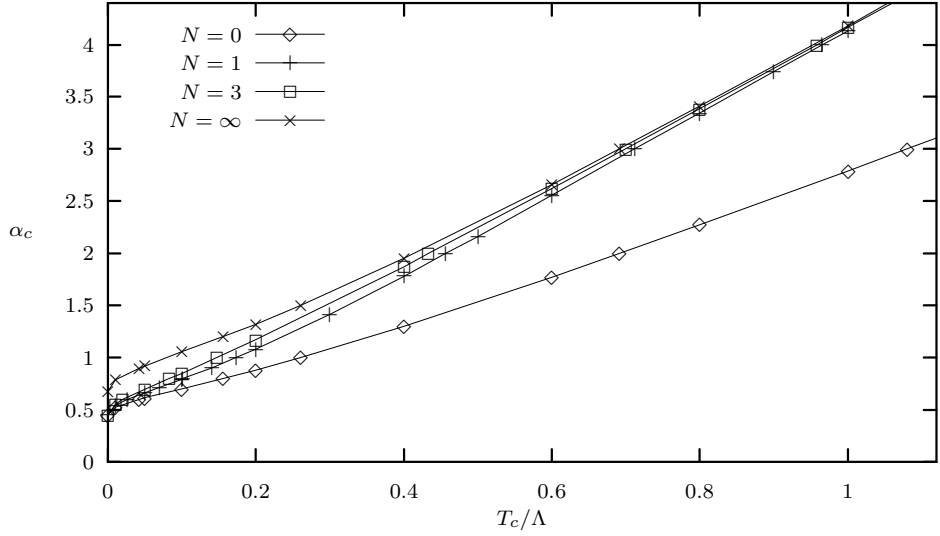
The critical curves for $N=0$, $N=1$, $N=3$ and $N=\infty$ are shown 
in Fig. 7.
Taking the $N\rightarrow\infty$ limit in Eq. (\ref{SD:IE:fin}) 
we find that the dynamical fermion mass for $N=\infty$ is obtained 
by replacing $\alpha$ by $(2/3)\alpha$ in the case $N=0$.
Thus the critical coupling constant $\alpha_{c}$ for $N=\infty$ is
given by $\alpha_c$ for $N=0$ multiplied by $2/3$. 
The longitudinal mode of the photon propagator acquires the 
non-vanishing thermal mass for $N\neq 0$.
Since the thermal photon mass $m_{ph}$ tends to suppress the 
gauge interaction, it has an effect of protecting the chiral symmetry. 
As is seen in Fig. 7 the effect of the thermal photon mass enhances 
the critical coupling $\alpha_c$ and becomes stronger as $T$ increases.

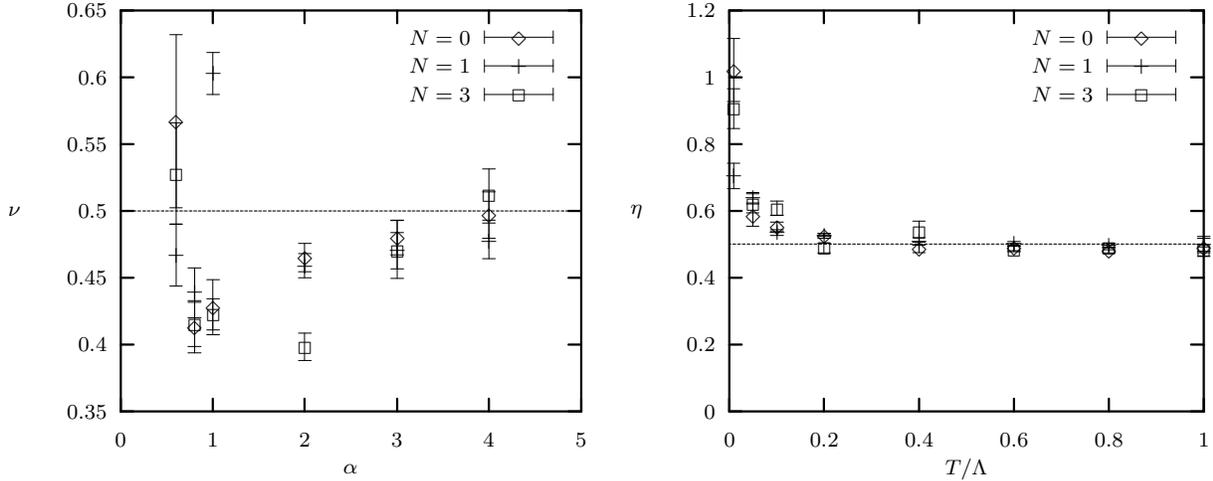
\begin{figure}
    \begin{minipage}{.5\linewidth}
    \begin{center}
\setlength{\unitlength}{0.240900pt}
\begin{picture}(1500,900)(0,0)
\footnotesize
\thicklines \path(177,135)(197,135)
\thicklines \path(900,135)(880,135)
\put(155,135){\makebox(0,0)[r]{0.35}}
\thicklines \path(177,240)(197,240)
\thicklines \path(900,240)(880,240)
\put(155,240){\makebox(0,0)[r]{0.4}}
\thicklines \path(177,345)(197,345)
\thicklines \path(900,345)(880,345)
\put(155,345){\makebox(0,0)[r]{0.45}}
\thicklines \path(177,450)(197,450)
\thicklines \path(900,450)(880,450)
\put(155,450){\makebox(0,0)[r]{0.5}}
\thicklines \path(177,555)(197,555)
\thicklines \path(900,555)(880,555)
\put(155,555){\makebox(0,0)[r]{0.55}}
\thicklines \path(177,660)(197,660)
\thicklines \path(900,660)(880,660)
\put(155,660){\makebox(0,0)[r]{0.6}}
\thicklines \path(177,765)(197,765)
\thicklines \path(900,765)(880,765)
\put(155,765){\makebox(0,0)[r]{0.65}}
\thicklines \path(177,135)(177,155)
\thicklines \path(177,765)(177,745)
\put(177,90){\makebox(0,0){0}}
\thicklines \path(322,135)(322,155)
\thicklines \path(322,765)(322,745)
\put(322,90){\makebox(0,0){1}}
\thicklines \path(466,135)(466,155)
\thicklines \path(466,765)(466,745)
\put(466,90){\makebox(0,0){2}}
\thicklines \path(611,135)(611,155)
\thicklines \path(611,765)(611,745)
\put(611,90){\makebox(0,0){3}}
\thicklines \path(755,135)(755,155)
\thicklines \path(755,765)(755,745)
\put(755,90){\makebox(0,0){4}}
\thicklines \path(900,135)(900,155)
\thicklines \path(900,765)(900,745)
\put(900,90){\makebox(0,0){5}}
\thicklines \path(177,135)(900,135)(900,765)(177,765)(177,135)
\put(0,450){\makebox(0,0)[l]{\shortstack{$\nu$}}}
\put(538,45){\makebox(0,0){$\alpha$}}
\put(726,723){\makebox(0,0)[r]{$N=0$}}
\thinlines \path(748,723)(856,723)
\thinlines \path(748,733)(748,713)
\thinlines \path(856,733)(856,713)
\thinlines \path(264,455)(264,727)
\thinlines \path(254,455)(274,455)
\thinlines \path(254,727)(274,727)
\thinlines \path(293,227)(293,309)
\thinlines \path(283,227)(303,227)
\thinlines \path(283,309)(303,309)
\thinlines \path(322,256)(322,342)
\thinlines \path(312,256)(332,256)
\thinlines \path(312,342)(332,342)
\thinlines \path(466,355)(466,399)
\thinlines \path(456,355)(476,355)
\thinlines \path(456,399)(476,399)
\thinlines \path(611,381)(611,436)
\thinlines \path(601,381)(621,381)
\thinlines \path(601,436)(621,436)
\thinlines \path(755,407)(755,480)
\thinlines \path(745,407)(765,407)
\thinlines \path(745,480)(765,480)
\put(264,591){\raisebox{-1.2pt}{\makebox(0,0){$\Diamond$}}}
\put(293,268){\raisebox{-1.2pt}{\makebox(0,0){$\Diamond$}}}
\put(322,299){\raisebox{-1.2pt}{\makebox(0,0){$\Diamond$}}}
\put(466,377){\raisebox{-1.2pt}{\makebox(0,0){$\Diamond$}}}
\put(611,408){\raisebox{-1.2pt}{\makebox(0,0){$\Diamond$}}}
\put(755,444){\raisebox{-1.2pt}{\makebox(0,0){$\Diamond$}}}
\put(802,723){\raisebox{-1.2pt}{\makebox(0,0){$\Diamond$}}}
\put(726,678){\makebox(0,0)[r]{$N=1$}}
\thinlines \path(748,678)(856,678)
\thinlines \path(748,688)(748,668)
\thinlines \path(856,688)(856,668)
\thinlines \path(264,332)(264,430)
\thinlines \path(254,332)(274,332)
\thinlines \path(254,430)(274,430)
\thinlines \path(293,282)(293,361)
\thinlines \path(283,282)(303,282)
\thinlines \path(283,361)(303,361)
\thinlines \path(322,633)(322,699)
\thinlines \path(312,633)(332,633)
\thinlines \path(312,699)(332,699)
\thinlines \path(466,345)(466,383)
\thinlines \path(456,345)(476,345)
\thinlines \path(456,383)(476,383)
\thinlines \path(611,344)(611,435)
\thinlines \path(601,344)(621,344)
\thinlines \path(601,435)(621,435)
\thinlines \path(755,375)(755,431)
\thinlines \path(745,375)(765,375)
\thinlines \path(745,431)(765,431)
\put(264,381){\makebox(0,0){$+$}}
\put(293,322){\makebox(0,0){$+$}}
\put(322,666){\makebox(0,0){$+$}}
\put(466,364){\makebox(0,0){$+$}}
\put(611,389){\makebox(0,0){$+$}}
\put(755,403){\makebox(0,0){$+$}}
\put(802,678){\makebox(0,0){$+$}}
\put(726,633){\makebox(0,0)[r]{$N=3$}}
\thinlines \path(748,633)(856,633)
\thinlines \path(748,643)(748,623)
\thinlines \path(856,643)(856,623)
\thinlines \path(264,429)(264,589)
\thinlines \path(254,429)(274,429)
\thinlines \path(254,589)(274,589)
\thinlines \path(293,237)(293,307)
\thinlines \path(283,237)(303,237)
\thinlines \path(283,307)(303,307)
\thinlines \path(322,263)(322,312)
\thinlines \path(312,263)(332,263)
\thinlines \path(312,312)(332,312)
\thinlines \path(466,215)(466,258)
\thinlines \path(456,215)(476,215)
\thinlines \path(456,258)(476,258)
\thinlines \path(611,359)(611,416)
\thinlines \path(601,359)(621,359)
\thinlines \path(601,416)(621,416)
\thinlines \path(755,436)(755,516)
\thinlines \path(745,436)(765,436)
\thinlines \path(745,516)(765,516)
\put(264,509){\raisebox{-1.2pt}{\makebox(0,0){$\Box$}}}
\put(293,272){\raisebox{-1.2pt}{\makebox(0,0){$\Box$}}}
\put(322,288){\raisebox{-1.2pt}{\makebox(0,0){$\Box$}}}
\put(466,236){\raisebox{-1.2pt}{\makebox(0,0){$\Box$}}}
\put(611,388){\raisebox{-1.2pt}{\makebox(0,0){$\Box$}}}
\put(755,476){\raisebox{-1.2pt}{\makebox(0,0){$\Box$}}}
\put(802,633){\raisebox{-1.2pt}{\makebox(0,0){$\Box$}}}
\thinlines \drawline[-50](177,450)(177,450)(184,450)(192,450)(199,450)
(206,450)(214,450)(221,450)(228,450)(235,450)(243,450)(250,450)(257,450)
(265,450)(272,450)(279,450)(287,450)(294,450)(301,450)(308,450)(316,450)
(323,450)(330,450)(338,450)(345,450)(352,450)(360,450)(367,450)(374,450)
(381,450)(389,450)(396,450)(403,450)(411,450)(418,450)(425,450)(433,450)
(440,450)(447,450)(455,450)(462,450)(469,450)(476,450)(484,450)(491,450)
(498,450)(506,450)(513,450)(520,450)(528,450)(535,450)
\thinlines \drawline[-50](535,450)(542,450)(549,450)(557,450)(564,450)
(571,450)(579,450)(586,450)(593,450)(601,450)(608,450)(615,450)(622,450)
(630,450)(637,450)(644,450)(652,450)(659,450)(666,450)(674,450)(681,450)
(688,450)(696,450)(703,450)(710,450)(717,450)(725,450)(732,450)(739,450)
(747,450)(754,450)(761,450)(769,450)(776,450)(783,450)(790,450)(798,450)
(805,450)(812,450)(820,450)(827,450)(834,450)(842,450)(849,450)(856,450)
(863,450)(871,450)(878,450)(885,450)(893,450)(900,450)
\end{picture}

    \end{center}
    \end{minipage}
\hfill
    \begin{minipage}{.6\linewidth}
    \begin{center}
\setlength{\unitlength}{0.240900pt}
\begin{picture}(1500,900)(0,0)
\footnotesize
\thicklines \path(155,135)(175,135)
\thicklines \path(900,135)(880,135)
\put(133,135){\makebox(0,0)[r]{0}}
\thicklines \path(155,240)(175,240)
\thicklines \path(900,240)(880,240)
\put(133,240){\makebox(0,0)[r]{0.2}}
\thicklines \path(155,345)(175,345)
\thicklines \path(900,345)(880,345)
\put(133,345){\makebox(0,0)[r]{0.4}}
\thicklines \path(155,450)(175,450)
\thicklines \path(900,450)(880,450)
\put(133,450){\makebox(0,0)[r]{0.6}}
\thicklines \path(155,555)(175,555)
\thicklines \path(900,555)(880,555)
\put(133,555){\makebox(0,0)[r]{0.8}}
\thicklines \path(155,660)(175,660)
\thicklines \path(900,660)(880,660)
\put(133,660){\makebox(0,0)[r]{1}}
\thicklines \path(155,765)(175,765)
\thicklines \path(900,765)(880,765)
\put(133,765){\makebox(0,0)[r]{1.2}}
\thicklines \path(155,135)(155,155)
\thicklines \path(155,765)(155,745)
\put(155,90){\makebox(0,0){0}}
\thicklines \path(304,135)(304,155)
\thicklines \path(304,765)(304,745)
\put(304,90){\makebox(0,0){0.2}}
\thicklines \path(453,135)(453,155)
\thicklines \path(453,765)(453,745)
\put(453,90){\makebox(0,0){0.4}}
\thicklines \path(602,135)(602,155)
\thicklines \path(602,765)(602,745)
\put(602,90){\makebox(0,0){0.6}}
\thicklines \path(751,135)(751,155)
\thicklines \path(751,765)(751,745)
\put(751,90){\makebox(0,0){0.8}}
\thicklines \path(900,135)(900,155)
\thicklines \path(900,765)(900,745)
\put(900,90){\makebox(0,0){1}}
\thicklines \path(155,135)(900,135)(900,765)(155,765)(155,135)
\put(0,450){\makebox(0,0)[l]{\shortstack{$\eta$}}}
\put(527,45){\makebox(0,0){$T/\Lambda$}}
\put(726,723){\makebox(0,0)[r]{$N=0$}}
\thinlines \path(748,723)(856,723)
\thinlines \path(748,733)(748,713)
\thinlines \path(856,733)(856,713)
\thinlines \path(162,622)(162,721)
\thinlines \path(152,622)(172,622)
\thinlines \path(152,721)(172,721)
\thinlines \path(192,426)(192,461)
\thinlines \path(182,426)(202,426)
\thinlines \path(182,461)(202,461)
\thinlines \path(230,418)(230,433)
\thinlines \path(220,418)(240,418)
\thinlines \path(220,433)(240,433)
\thinlines \path(304,407)(304,415)
\thinlines \path(294,407)(314,407)
\thinlines \path(294,415)(314,415)
\thinlines \path(453,385)(453,397)
\thinlines \path(443,385)(463,385)
\thinlines \path(443,397)(463,397)
\thinlines \path(602,388)(602,395)
\thinlines \path(592,388)(612,388)
\thinlines \path(592,395)(612,395)
\thinlines \path(751,384)(751,393)
\thinlines \path(741,384)(761,384)
\thinlines \path(741,393)(761,393)
\thinlines \path(900,379)(900,407)
\thinlines \path(890,379)(910,379)
\thinlines \path(890,407)(910,407)
\put(162,671){\raisebox{-1.2pt}{\makebox(0,0){$\Diamond$}}}
\put(192,443){\raisebox{-1.2pt}{\makebox(0,0){$\Diamond$}}}
\put(230,425){\raisebox{-1.2pt}{\makebox(0,0){$\Diamond$}}}
\put(304,411){\raisebox{-1.2pt}{\makebox(0,0){$\Diamond$}}}
\put(453,391){\raisebox{-1.2pt}{\makebox(0,0){$\Diamond$}}}
\put(602,392){\raisebox{-1.2pt}{\makebox(0,0){$\Diamond$}}}
\put(751,388){\raisebox{-1.2pt}{\makebox(0,0){$\Diamond$}}}
\put(900,393){\raisebox{-1.2pt}{\makebox(0,0){$\Diamond$}}}
\put(802,723){\raisebox{-1.2pt}{\makebox(0,0){$\Diamond$}}}
\put(726,678){\makebox(0,0)[r]{$N=1$}}
\thinlines \path(748,678)(856,678)
\thinlines \path(748,688)(748,668)
\thinlines \path(856,688)(856,668)
\thinlines \path(162,485)(162,525)
\thinlines \path(152,485)(172,485)
\thinlines \path(152,525)(172,525)
\thinlines \path(192,463)(192,479)
\thinlines \path(182,463)(202,463)
\thinlines \path(182,479)(202,479)
\thinlines \path(230,412)(230,421)
\thinlines \path(220,412)(240,412)
\thinlines \path(220,421)(240,421)
\thinlines \path(304,410)(304,415)
\thinlines \path(294,410)(314,410)
\thinlines \path(294,415)(314,415)
\thinlines \path(453,397)(453,408)
\thinlines \path(443,397)(463,397)
\thinlines \path(443,408)(463,408)
\thinlines \path(602,395)(602,402)
\thinlines \path(592,395)(612,395)
\thinlines \path(592,402)(612,402)
\thinlines \path(751,392)(751,399)
\thinlines \path(741,392)(761,392)
\thinlines \path(741,399)(761,399)
\thinlines \path(900,387)(900,410)
\thinlines \path(890,387)(910,387)
\thinlines \path(890,410)(910,410)
\put(162,505){\makebox(0,0){$+$}}
\put(192,471){\makebox(0,0){$+$}}
\put(230,417){\makebox(0,0){$+$}}
\put(304,412){\makebox(0,0){$+$}}
\put(453,403){\makebox(0,0){$+$}}
\put(602,399){\makebox(0,0){$+$}}
\put(751,396){\makebox(0,0){$+$}}
\put(900,398){\makebox(0,0){$+$}}
\put(802,678){\makebox(0,0){$+$}}
\put(726,633){\makebox(0,0)[r]{$N=3$}}
\thinlines \path(748,633)(856,633)
\thinlines \path(748,643)(748,623)
\thinlines \path(856,643)(856,623)
\thinlines \path(162,580)(162,642)
\thinlines \path(152,580)(172,580)
\thinlines \path(152,642)(172,642)
\thinlines \path(192,447)(192,477)
\thinlines \path(182,447)(202,447)
\thinlines \path(182,477)(202,477)
\thinlines \path(230,443)(230,466)
\thinlines \path(220,443)(240,443)
\thinlines \path(220,466)(240,466)
\thinlines \path(304,384)(304,400)
\thinlines \path(294,384)(314,384)
\thinlines \path(294,400)(314,400)
\thinlines \path(453,401)(453,434)
\thinlines \path(443,401)(463,401)
\thinlines \path(443,434)(463,434)
\thinlines \path(602,386)(602,395)
\thinlines \path(592,386)(612,386)
\thinlines \path(592,395)(612,395)
\thinlines \path(751,389)(751,397)
\thinlines \path(741,389)(761,389)
\thinlines \path(741,397)(761,397)
\thinlines \path(900,385)(900,394)
\thinlines \path(890,385)(910,385)
\thinlines \path(890,394)(910,394)
\put(162,611){\raisebox{-1.2pt}{\makebox(0,0){$\Box$}}}
\put(192,462){\raisebox{-1.2pt}{\makebox(0,0){$\Box$}}}
\put(230,454){\raisebox{-1.2pt}{\makebox(0,0){$\Box$}}}
\put(304,392){\raisebox{-1.2pt}{\makebox(0,0){$\Box$}}}
\put(453,418){\raisebox{-1.2pt}{\makebox(0,0){$\Box$}}}
\put(602,390){\raisebox{-1.2pt}{\makebox(0,0){$\Box$}}}
\put(751,393){\raisebox{-1.2pt}{\makebox(0,0){$\Box$}}}
\put(900,389){\raisebox{-1.2pt}{\makebox(0,0){$\Box$}}}
\put(802,633){\raisebox{-1.2pt}{\makebox(0,0){$\Box$}}}
\thinlines \drawline[-50](155,398)(155,398)(163,398)(170,398)(178,398)
(185,398)(193,398)(200,398)(208,398)(215,398)(223,398)(230,398)(238,398)
(245,398)(253,398)(260,398)(268,398)(275,398)(283,398)(290,398)(298,398)
(306,398)(313,398)(321,398)(328,398)(336,398)(343,398)(351,398)(358,398)
(366,398)(373,398)(381,398)(388,398)(396,398)(403,398)(411,398)(418,398)
(426,398)(433,398)(441,398)(448,398)(456,398)(464,398)(471,398)(479,398)
(486,398)(494,398)(501,398)(509,398)(516,398)(524,398)
\thinlines \drawline[-50](524,398)(531,398)(539,398)(546,398)(554,398)
(561,398)(569,398)(576,398)(584,398)(591,398)(599,398)(607,398)(614,398)
(622,398)(629,398)(637,398)(644,398)(652,398)(659,398)(667,398)(674,398)
(682,398)(689,398)(697,398)(704,398)(712,398)(719,398)(727,398)(734,398)
(742,398)(749,398)(757,398)(765,398)(772,398)(780,398)(787,398)(795,398)
(802,398)(810,398)(817,398)(825,398)(832,398)(840,398)(847,398)(855,398)
(862,398)(870,398)(877,398)(885,398)(892,398)(900,398)
\end{picture}

    \end{center}
    \end{minipage}
    \caption{Critical exponents $\nu$ and $\eta$ estimated 
    at $p=0.1\Lambda$.}
\end{figure}
In Fig. 8 the critical exponents $\nu$ and $\eta$ are shown as a
function of the temperature and coupling constant respectively.
As is seen in Fig. 8 the critical exponents do not depend on 
the number of fermion flavors $N$ guaranteeing the accuracy of our numerical 
analysis.
Taking the average of the critical exponents $\nu$ for $\alpha=1, 2, 3, 4$ 
and $\eta$ for $T/\Lambda=0.2, 0.4, 0.6, 0.8, 1.0$ we obtain
\begin{equation}
\nu \sim \left\{
\begin{array}{ccc}
 0.46
&\mbox{    for  }& N=0 ,\\
 0.50
&\mbox{    for  }& N=1 ,\\
 0.45
&\mbox{    for  }& N=3 ,
\end{array}
\right.
\end{equation}
and
\begin{equation}
\eta \sim \left\{
\begin{array}{ccc}
 0.49
&\mbox{    for  }& N=0 ,\\
 0.51
&\mbox{    for  }& N=1 ,\\
 0.49
&\mbox{    for  }& N=3 ,
\end{array}
\right.
\end{equation}
respectively. Therefore our results are consistent with 
$\nu=1/2$ and $\eta=1/2$ which are in accord with the 
critical exponents in the four-fermion theory \cite{IKM}.

In Fig. 9 the $N$ dependence of the mass function is presented.
Since the longitudinal mode for photon propagator acquires
the thermal mass for  $N \neq 0$, the photon propagator has
only one massless mode.
On the other hand the photon propagator has three massless modes
for $N=0$.
As is seen in Fig. 9 the dynamical fermion mass strongly
depends on the number of massless modes in the photon propagator
near the critical temperature where the instantaneous exchange
approximation is valid.
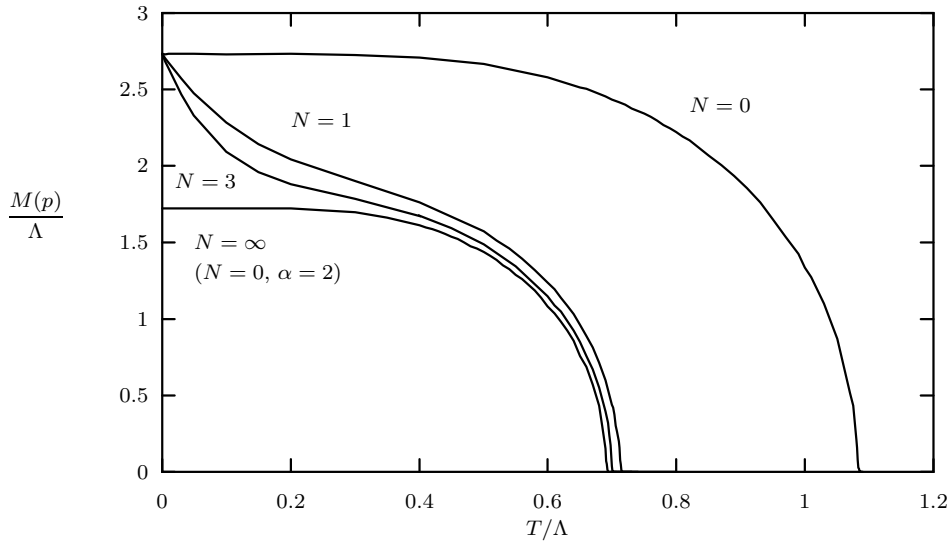
\begin{figure}
\begin{center}
\setlength{\unitlength}{0.240900pt}
\begin{picture}(1500,900)(0,0)
\footnotesize
\thicklines \path(245,135)(265,135)
\thicklines \path(1456,135)(1436,135)
\put(223,135){\makebox(0,0)[r]{0}}
\thicklines \path(245,255)(265,255)
\thicklines \path(1456,255)(1436,255)
\put(223,255){\makebox(0,0)[r]{0.5}}
\thicklines \path(245,375)(265,375)
\thicklines \path(1456,375)(1436,375)
\put(223,375){\makebox(0,0)[r]{1}}
\thicklines \path(245,496)(265,496)
\thicklines \path(1456,496)(1436,496)
\put(223,496){\makebox(0,0)[r]{1.5}}
\thicklines \path(245,616)(265,616)
\thicklines \path(1456,616)(1436,616)
\put(223,616){\makebox(0,0)[r]{2}}
\thicklines \path(245,736)(265,736)
\thicklines \path(1456,736)(1436,736)
\put(223,736){\makebox(0,0)[r]{2.5}}
\thicklines \path(245,856)(265,856)
\thicklines \path(1456,856)(1436,856)
\put(223,856){\makebox(0,0)[r]{3}}
\thicklines \path(245,135)(245,155)
\thicklines \path(245,856)(245,836)
\put(245,90){\makebox(0,0){0}}
\thicklines \path(447,135)(447,155)
\thicklines \path(447,856)(447,836)
\put(447,90){\makebox(0,0){0.2}}
\thicklines \path(649,135)(649,155)
\thicklines \path(649,856)(649,836)
\put(649,90){\makebox(0,0){0.4}}
\thicklines \path(850,135)(850,155)
\thicklines \path(850,856)(850,836)
\put(850,90){\makebox(0,0){0.6}}
\thicklines \path(1052,135)(1052,155)
\thicklines \path(1052,856)(1052,836)
\put(1052,90){\makebox(0,0){0.8}}
\thicklines \path(1254,135)(1254,155)
\thicklines \path(1254,856)(1254,836)
\put(1254,90){\makebox(0,0){1}}
\thicklines \path(1456,135)(1456,155)
\thicklines \path(1456,856)(1456,836)
\put(1456,90){\makebox(0,0){1.2}}
\thicklines \path(245,135)(1456,135)(1456,856)(245,856)(245,135)
\put(0,539){\makebox(0,0)[l]
{\shortstack{$\displaystyle\frac{M(p)}{\Lambda}$}}}
\put(850,45){\makebox(0,0){$T/\Lambda$}}
\put(1073,712){\makebox(0,0)[l]{$N=0$}}
\put(447,688){\makebox(0,0)[l]{$N=1$}}
\put(265,592){\makebox(0,0)[l]{$N=3$}}
\put(295,496){\makebox(0,0)[l]{$N=\infty$}}
\put(295,447){\makebox(0,0)[l]{($N=0$, $\alpha=2$)}}
\thicklines \path(245,792)(245,792)(250,791)(255,792)(295,792)(346,791)
(447,792)(548,790)(649,786)(750,776)(850,755)(901,739)(911,737)(921,733)
(931,729)(941,725)(951,720)(962,716)(972,712)(982,706)(992,702)(1002,698)
(1012,693)(1022,687)(1032,680)(1042,675)(1052,669)(1062,662)(1073,656)
(1083,648)(1103,632)(1123,617)(1143,600)(1163,581)(1184,559)(1204,532)
(1224,504)(1244,478)(1254,456)(1264,441)(1284,400)(1305,344)(1325,259)
(1330,239)(1335,185)(1336,173)(1337,154)(1338,142)(1339,140)(1340,136)
(1341,136)
\thicklines \path(1341,136)(1345,135)
\thicklines \path(245,549)(245,549)(250,549)(255,549)(295,549)(346,549)
(396,549)(447,549)(497,546)(548,543)(598,535)(649,523)(659,519)(669,517)
(679,513)(689,509)(699,505)(709,501)(719,496)(729,490)(739,486)(750,481)
(760,475)(770,468)(780,460)(790,454)(800,445)(810,438)(820,429)(830,420)
(840,409)(850,395)(861,384)(871,371)(881,357)(891,341)(901,317)(911,300)
(921,272)(931,239)(936,206)(941,172)(942,152)(943,148)(944,139)(945,136)
(946,135)(947,135)(949,135)(951,135)(962,135)
\thicklines \path(962,135)(972,135)(982,135)(992,135)
\thicklines \path(245,792)(245,792)(250,785)(255,778)(275,753)(295,730)
(346,684)(346,684)(396,650)(447,626)(548,592)(649,559)(750,513)(770,499)
(790,486)(810,470)(830,452)(850,433)(861,422)(871,409)(881,396)(891,383)
(901,366)(911,349)(921,331)(931,307)(941,279)(951,241)(953,236)(955,225)
(957,208)(959,199)(962,183)(963,173)(964,154)(965,147)(966,137)(968,135)
(972,135)(982,135)
\thicklines \path(245,792)(245,792)(246,789)(250,780)(255,769)(275,728)
(295,695)(346,638)(396,606)(447,587)(548,564)(649,537)(649,538)(699,518)
(750,493)(800,458)(850,411)(850,411)(861,397)(871,387)(881,372)(891,357)
(901,339)(911,317)(921,296)(931,268)(941,230)(945,212)(947,193)(949,177)
(951,144)(952,136)(962,135)(972,135)(972,135)(982,135)(992,135)(992,135)
(1002,135)(1012,135)(1012,135)(1022,135)(1032,135)(1032,135)(1042,135)
(1052,135)(1052,135)
\end{picture}

\caption{Dynamical fermion mass for $N=0, N=1, N=3$ and $N=\infty$
as a function of the temperature $T$ at $\alpha=3$ and $p=0.1\Lambda$.}
\end{center}
\end{figure}

Note that the behaviors of the critical curve and the dynamical fermion 
mass for finite $N$ seem different from the results in Ref. \cite{KY}.
In Ref. \cite{KY} the thermal photon mass is introduced both for the 
longitudinal and transversal mode. 
In this case there is no massless mode for the photon propagator.

\section{Conclusion}

We have investigated the mechanism of the chiral symmetry breaking in
the strong-coupling Abelian gauge theories at finite temperature.  The
Schwinger-Dyson equation in Landau gauge is solved numerically within
the framework of the instantaneous exchange approximation which is
valid at high temperature.
The effect of the hard thermal loop for the photon propagator on the
phase transition has been also clarified.
Here are our results in order:
\begin{enumerate}
 \item The chiral phase transition is found to be of the 2nd order in
       the high temperature region. Thus the physical mass
       obeys the scaling law of the mean-field type. The effect of
       temperature on the chiral phase transition seems to be equivalent
       to the fermion loop effect in strong coupling QED at $T = 0$. 
 \item The critical temperature grows linearly as the critical coupling
       constant grows in the high temperature region. This result is
       opposed to that of Ref. \cite{KY}
       where they claimed that the chiral symmetry is always restored at
       finite temperature no matter how large the coupling constant is
       taken.
       The origin of this discrepancy lies in the vacuum polarization
       function. We have taken its transverse and longitudinal parts
       separately (see Eq. (\ref{VP})) while in Ref. \cite{KY}
       they were assumed to be the same.  Note that our choice of the
       function gives no constraints on the number of fermion flavors,
       whereas in strong coupling QED at $T = 0$ there might be the
       critical value $N_c$ above which there is the chiral symmetric
       phase only \cite{KN}.
 \item The thermal photon mass reduces the effect of electromagnetic
       interaction. In the limit of the infinite number of fermion
       flavors the behavior of the mass function in the case with the
       thermal photon mass ($N = \infty$) is the same as in the case without
       the thermal photon mass ($N = 0$) with the coupling
       constant $\alpha$ replaced by $2\alpha /3$.
\end{enumerate}

The next step that we should put forward is to extend our analysis
to the one in the low temperature region. Along this line there are
some difficulties to be overcome:
\begin{enumerate}
 \item The vacuum polarization function at low temperature has to be
       estimated with some suitable approximation or in numerical calculations.
 \item Proper approximations have to be found out in order to solve
       the Schwinger-Dyson equation at low temperature otherwise the
       equation is much complicated.
\end{enumerate}
We expect that the structure of the chiral phase transition is the same
as in the case we have investigated.
This will be a subject in the forthcoming paper.

\section*{Acknowledgements}

The authors would like to thank S. D. Odintsov for useful
conversations in the early stage of the present work and K.-I. Kondo
for enlightening discussions.
The present work is performed under the auspice of Monbusho Fund
(Grant-in-Aid for Scientific Research (C) and Grant-in-Aid for
Encouragement of Young Scientists from the Ministry of Education,
Science and Culture) with contract numbers 08640377(T.M.), 11640280(T.M.)
and 11740154(T.I.) respectively.

\appendix
\section*{Appendix}
\setcounter{equation}{0}
\makeatletter
\def\theequation
   {A.\arabic{equation}}
\makeatother

Here we argue the critical value of coupling constant $\alpha_c$ at
vanishing temperature within the instantaneous exchange approximation,
though this approximation is valid only at high temperature.

We begin with the linearized Schwinger-Dyson equation
\begin{equation}
 M(p) = \frac{3\alpha}{4\pi p} \int_0^\Lambda dq M(q) 
  \ln\frac{(p+q)^2}{(p-q)^2} ,
\label{eq:A-SDeq}
\end{equation}
which is taken from 
Eq.(\ref{SD:IE:fin})
by setting $T = 0$ and neglecting the mass function in the denominator.
If we take only the first term in the expansion of the logarithm,
\begin{equation}
 \ln\frac{(p+q)^2}{(p-q)^2} = \sum_{n = 0}^\infty \frac{4}{2n + 1}
  \left[ \left( \frac{q}{p}\right)^{2n + 1} \theta(p-q)
   + \left( \frac{p}{q}\right)^{2n + 1} \theta(q-p) \right] ,
\label{eq:A-logarithm}
\end{equation}
it is easily found
that the Schwinger-Dyson equation,
\begin{equation}
 M(p) \simeq \frac{3\alpha}{\pi p} \int_0^\Lambda dq M(q) 
  \left[ \frac{q}{p} \theta(p-q) + \frac{p}{q} \theta(q-p) \right] ,
\end{equation}
has the following nontrivial solution:
\begin{eqnarray}
\hfill && M(p) \propto p^\lambda , \\
 && \hspace*{10mm}\lambda = \left\{
        \begin{array}{lc}
         \displaystyle -1 + \sqrt{1-\frac{6\alpha}{\pi}} 
          & \left(\displaystyle 0 < \alpha < \frac{\pi}{6}\right), \\
         & \\
         \displaystyle -1 \pm i\sqrt{\frac{6\alpha}{\pi}-1} 
          & \left(\displaystyle \alpha > \frac{\pi}{6}\right). \\
        \end{array}
        \right .
\end{eqnarray}
Note that, when the value of coupling constant $\alpha$ lies in the region
$\displaystyle \alpha < \pi/6$, $\lambda$ is real and the system
is in the symmetric phase, while the value of $\alpha$ lies in the region
$\displaystyle \alpha > \pi/6$, $\lambda$ is complex so that the
solution of the Schwinger-Dyson equation is of the oscillatory function
and the system is in the broken phase \cite{FK}.
The critical value $\alpha_c$ in this case is $\displaystyle
\pi/6$, which is half of the well known value $\displaystyle
\pi/3$ in Landau gauge, because of the instantaneous exchange
approximation and the simplification of the logarithm.

The evaluation of the critical value in the case including the full term 
of Eq.(\ref{eq:A-logarithm}) is made as follows.
The Schwinger-Dyson equation (\ref{eq:A-SDeq}) with the expansion
Eq.(\ref{eq:A-logarithm}) reads
\begin{eqnarray}
  M(p) = \frac{3\alpha}{\pi p} \sum_{n = 0}^\infty \frac{1}{2n + 1}
  \int_0^\Lambda dq M(q)
  \left[ \left( \frac{q}{p}\right)^{2n + 1} \theta(p-q)
   + \left( \frac{p}{q}\right)^{2n + 1} \theta(q-p) \right] ,
\label{eq:A-SDeqInExpansion}
\end{eqnarray}
and we take the solution of the form $\displaystyle M(p) \propto
p^\lambda$, which is valid in the above simplified case.
Note that the value of $\lambda$ is restricted 
by $ -2 < {\rm Re} \lambda < 0 $
due to the condition that the integral in
Eq.(\ref{eq:A-SDeqInExpansion}) should be finite.
Then the Schwinger-Dyson equation reduces to
\begin{equation}
 \frac{3\alpha}{2(\lambda + 1)} \tan\frac{\pi(\lambda + 1)}{2} = 1 ,
\label{eq:A-lambda}
\end{equation}
by performing the integral and using the formula 
\begin{equation} 
  \sum_{n = 0}^\infty
  \frac{1}{(2n + 1)^2 - x^2} = \frac{\pi}{4x}\tan\frac{\pi x}{2} .
\end{equation} 
The real solution of Eq.(\ref{eq:A-lambda}) is easily found by considering
the inclinations of $\tan \pi(\lambda + 1) / 2$ and $2(\lambda + 1) / 3$
at $\lambda = -1$.  The result is that the nontrivial solution $\lambda$
is real when $\displaystyle 0 < \alpha < 4/(3\pi)$ which corresponds to
the symmetric phase, whereas Eq.(\ref{eq:A-lambda}) has complex
solutions when $\displaystyle \alpha > 4/(3\pi)$ which corresponds to
the broken phase.  Hence the critical value of coupling constant in this
case is 
\begin{equation}
\alpha_c = \frac{4}{3\pi}
\end{equation}

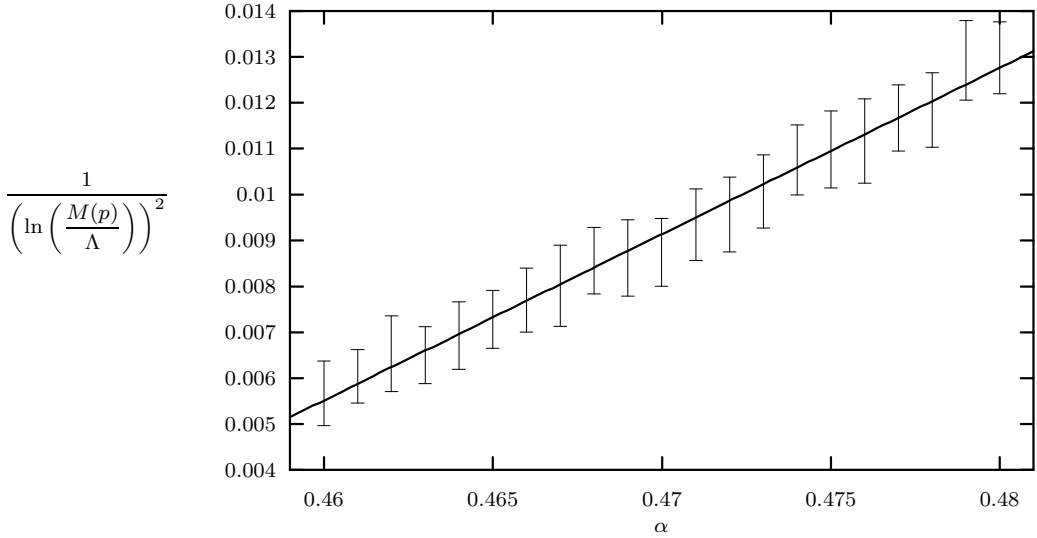
\begin{figure}
\begin{center}
\setlength{\unitlength}{0.240900pt}
\begin{picture}(1500,900)(0,0)
\footnotesize
\thicklines \path(289,135)(309,135)
\thicklines \path(1456,135)(1436,135)
\put(267,135){\makebox(0,0)[r]{0.004}}
\thicklines \path(289,207)(309,207)
\thicklines \path(1456,207)(1436,207)
\put(267,207){\makebox(0,0)[r]{0.005}}
\thicklines \path(289,279)(309,279)
\thicklines \path(1456,279)(1436,279)
\put(267,279){\makebox(0,0)[r]{0.006}}
\thicklines \path(289,351)(309,351)
\thicklines \path(1456,351)(1436,351)
\put(267,351){\makebox(0,0)[r]{0.007}}
\thicklines \path(289,423)(309,423)
\thicklines \path(1456,423)(1436,423)
\put(267,423){\makebox(0,0)[r]{0.008}}
\thicklines \path(289,496)(309,496)
\thicklines \path(1456,496)(1436,496)
\put(267,496){\makebox(0,0)[r]{0.009}}
\thicklines \path(289,568)(309,568)
\thicklines \path(1456,568)(1436,568)
\put(267,568){\makebox(0,0)[r]{0.01}}
\thicklines \path(289,640)(309,640)
\thicklines \path(1456,640)(1436,640)
\put(267,640){\makebox(0,0)[r]{0.011}}
\thicklines \path(289,712)(309,712)
\thicklines \path(1456,712)(1436,712)
\put(267,712){\makebox(0,0)[r]{0.012}}
\thicklines \path(289,784)(309,784)
\thicklines \path(1456,784)(1436,784)
\put(267,784){\makebox(0,0)[r]{0.013}}
\thicklines \path(289,856)(309,856)
\thicklines \path(1456,856)(1436,856)
\put(267,856){\makebox(0,0)[r]{0.014}}
\thicklines \path(342,135)(342,155)
\thicklines \path(342,856)(342,836)
\put(342,90){\makebox(0,0){0.46}}
\thicklines \path(607,135)(607,155)
\thicklines \path(607,856)(607,836)
\put(607,90){\makebox(0,0){0.465}}
\thicklines \path(873,135)(873,155)
\thicklines \path(873,856)(873,836)
\put(873,90){\makebox(0,0){0.47}}
\thicklines \path(1138,135)(1138,155)
\thicklines \path(1138,856)(1138,836)
\put(1138,90){\makebox(0,0){0.475}}
\thicklines \path(1403,135)(1403,155)
\thicklines \path(1403,856)(1403,836)
\put(1403,90){\makebox(0,0){0.48}}
\thicklines \path(289,135)(1456,135)(1456,856)(289,856)(289,135)
\put(-160,539){\makebox(0,0)[l]
{\shortstack{$\displaystyle\frac{1}
{\left( \ln\left( \displaystyle\frac{M(p)}{\Lambda}\right)\right)^2}$}}}
\put(872,45){\makebox(0,0){{$\alpha$}}}
\thinlines \path(342,306)(342,205)
\thinlines \path(332,306)(352,306)
\thinlines \path(332,205)(352,205)
\thinlines \path(395,324)(395,240)
\thinlines \path(385,324)(405,324)
\thinlines \path(385,240)(405,240)
\thinlines \path(448,377)(448,258)
\thinlines \path(438,377)(458,377)
\thinlines \path(438,258)(458,258)
\thinlines \path(501,360)(501,271)
\thinlines \path(491,360)(511,360)
\thinlines \path(491,271)(511,271)
\thinlines \path(554,399)(554,293)
\thinlines \path(544,399)(564,399)
\thinlines \path(544,293)(564,293)
\thinlines \path(607,417)(607,326)
\thinlines \path(597,417)(617,417)
\thinlines \path(597,326)(617,326)
\thinlines \path(660,452)(660,352)
\thinlines \path(650,452)(670,452)
\thinlines \path(650,352)(670,352)
\thinlines \path(713,488)(713,361)
\thinlines \path(703,488)(723,488)
\thinlines \path(703,361)(723,361)
\thinlines \path(766,516)(766,412)
\thinlines \path(756,516)(776,516)
\thinlines \path(756,412)(776,412)
\thinlines \path(819,528)(819,408)
\thinlines \path(809,528)(829,528)
\thinlines \path(809,408)(829,408)
\thinlines \path(872,530)(872,424)
\thinlines \path(862,530)(882,530)
\thinlines \path(862,424)(882,424)
\thinlines \path(926,577)(926,464)
\thinlines \path(916,577)(936,577)
\thinlines \path(916,464)(936,464)
\thinlines \path(979,595)(979,478)
\thinlines \path(969,595)(989,595)
\thinlines \path(969,478)(989,478)
\thinlines \path(1032,630)(1032,515)
\thinlines \path(1022,630)(1042,630)
\thinlines \path(1022,515)(1042,515)
\thinlines \path(1085,677)(1085,567)
\thinlines \path(1075,677)(1095,677)
\thinlines \path(1075,567)(1095,567)
\thinlines \path(1138,699)(1138,578)
\thinlines \path(1128,699)(1148,699)
\thinlines \path(1128,578)(1148,578)
\thinlines \path(1191,718)(1191,586)
\thinlines \path(1181,718)(1201,718)
\thinlines \path(1181,586)(1201,586)
\thinlines \path(1244,740)(1244,636)
\thinlines \path(1234,740)(1254,740)
\thinlines \path(1234,636)(1254,636)
\thinlines \path(1297,759)(1297,642)
\thinlines \path(1287,759)(1307,759)
\thinlines \path(1287,642)(1307,642)
\thinlines \path(1350,841)(1350,716)
\thinlines \path(1340,841)(1360,841)
\thinlines \path(1340,716)(1360,716)
\thinlines \path(1403,839)(1403,726)
\thinlines \path(1393,839)(1413,839)
\thinlines \path(1393,726)(1413,726)
\thicklines \path(289,218)(289,218)(301,224)(313,230)(324,236)(336,241)
(348,247)(360,253)(372,259)(383,265)(395,270)(407,276)(419,282)(430,288)
(442,294)(454,299)(466,305)(478,311)(489,317)(501,323)(513,328)(525,334)
(537,340)(548,346)(560,352)(572,357)(584,363)(595,369)(607,375)(619,381)
(631,386)(643,392)(654,398)(666,404)(678,410)(690,416)(702,421)(713,427)
(725,433)(737,439)(749,445)(761,450)(772,456)(784,462)(796,468)(808,474)
(819,479)(831,485)(843,491)(855,497)(867,503)
\thicklines \path(867,503)(878,508)(890,514)(902,520)(914,526)(926,532)
(937,537)(949,543)(961,549)(973,555)(984,561)(996,566)(1008,572)
(1020,578)(1032,584)(1043,590)(1055,595)(1067,601)(1079,607)(1091,613)
(1102,619)(1114,624)(1126,630)(1138,636)(1150,642)(1161,648)(1173,653)
(1185,659)(1197,665)(1208,671)(1220,677)(1232,682)(1244,688)(1256,694)
(1267,700)(1279,706)(1291,711)(1303,717)(1315,723)(1326,729)(1338,735)
(1350,740)(1362,746)(1373,752)(1385,758)(1397,764)(1409,770)(1421,775)
(1432,781)(1444,787)(1456,793)
\end{picture}

\caption{$1/( \ln(M(p)/\Lambda) )^2$ vs. $\alpha$ with a fitted line. The
critical value is given by $\alpha_c = 0.44477 \pm 0.00053$.}
\end{center}
\end{figure}
The exact value of critical coupling constant can be evaluated through
numerical calculation, by fitting data with the function in the form of
Miransky scaling law, and we find that the result is $\alpha_c \approx
0.445$. 
The numerical value 0.445 is in discrepancy with our theoretical
prediction $4/(3\pi)$. It is easy to understand that this discrepancy is 
due to the fact that we made a linearization approximation.
One should remember that the mass function in the denominator in 
Eq.(\ref{SD:IE:fin})
is neglected in Eq.(\ref{eq:A-SDeq}). This means that the exact value
of $\alpha_c$ must be slightly larger than $\displaystyle
4/(3\pi)$ and would be in the range
\begin{equation}
 \frac{4}{3\pi} < \alpha_c \lesssim \frac{\pi}{6}.
\end{equation}
It is easy to see that our numerical result satisfies this speculation.

\end{document}